\documentclass[12pt,a4paper]{JHEP3}
\usepackage{graphicx}
\def\be{\begin{equation}}
\def\ee{\end{equation}}
\def\bea{\begin{eqnarray}}
\def\eea{\end{eqnarray}}

\def\lx{\left}
\def\rx{\right}
\def\la{\langle}
\def\ra{\rangle}

\def\a{\alpha}
\def\b{\beta}
\def\g{\gamma}
\def\d{\delta}
\def\l{\lambda}
\def\r{\rho}
\def\s{\sigma}
\def\t{\tau}
\def\w{\omega}

\def\idt{\int_0^1  \!\!\! d\t\ }
\def\ids{\int_0^1 \!\!\! d\s\ }  
\def\idts{\int_0^1 \!\!\! d\t\! \int_0^1  \!\!\! d\s \ }
\def\idtt{\int_0^1 \!\!\! d\t_1\! \int_0^1  \!\!\! d\t_2\ }
\def\idttt{\int_0^1 \!\!\! d\t_1\! \int_0^1  \!\!\! d\t_2\! 
           \int_0^1 \!\!\! d\t_3\ }

\def\del{\Delta}
\def\ddel{{}^\bullet\! \Delta}

\def\ddeld{{}^{\bullet}\! \Delta^{\hskip -.5mm \bullet}}

\def\B{{\cal B}}
\def\dB{{}^\bullet\! {\cal B}}
\def\Bd{{\cal B}^\bullet}
\def\dBd{{}^{\bullet}\! {\cal B}^\bullet}

\def\F{{\cal F}}
\def\dF{{}^\bullet\! {\cal F}}
\def\Fd{{\cal F}^\bullet}
\def\dFd{{}^{\bullet}\! {\cal F}^\bullet}

\def\Dgh{\Delta_{gh}}

\def\sqr#1#2{{\vcenter{\vbox{\hrule height.#2pt
     \hbox{\vrule width.#2pt height#1pt \kern#1pt
           \vrule width.#2pt}
       \hrule height.#2pt}}}}
\def\Box{\mathchoice\sqr55\sqr55\sqr{2.1}3\sqr{1.5}3}
\def\Tr{{\rm Tr}~}

\title{BRST treatment of zero modes for the worldline formalism 
in curved space}

\author{Fiorenzo Bastianelli, Olindo Corradini and Andrea Zirotti
\\Dipartimento  di Fisica, Universit{\`a} di Bologna 
and  INFN, Sezione di Bologna\\
Via Irnerio 46, I-40126 Bologna, Italy 
\\ E-mail: \email{bastianelli@bo.infn.it}, \email{corradini@bo.infn.it},\\
\hskip1.45cm \email{zirotti@bo.infn.it}}

\abstract{
One-loop quantities in QFT can be computed in an efficient way using the 
worldline formalism. The latter rests on the ability of calculating
1D path integrals on the circle. In this paper we give a systematic 
discussion for treating zero modes on the circle of 1D path integrals
for both bosonic and supersymmetric nonlinear sigma models, following 
an approach originally introduced by Friedan. We use BRST techniques and 
place a special emphasis on the issue of reparametrization invariance. 
Various examples are extensively analyzed to verify and test the general 
set-up. In particular, we explicitly check that the chiral anomaly, which 
can be obtained by the semiclassical approximation of a supersymmetric 1D 
path integral, does not receive higher order worldline contributions, 
as implied by supersymmetry.
}

\preprint{hep-th/0312064}

\begin{document}

\section{Introduction}

The worldline formalism is an efficient and economical way of calculating  
Feynman diagrams~\cite{Schubert:2001he}.
It describes the propagation of various relativistic
particles by one dimensional (1D) path integrals. 
Recently we have shown how to extend this method to the case of spin 0 
and spin 1/2 particles coupled to background gravity
\cite{Bastianelli:2002fv,Bastianelli:2002qw}.
Other applications run from the calculation of the heat kernel 
to the calculation of chiral and trace anomalies.  

In all these cases, the 1D path integrals are calculated on the circle
with a finite propagation time, the proper time. In many circumstances
this proper time is integrated over, as it represents the only modulus of the 
circle. The actions in the path integrals are those of one dimensional 
nonlinear sigma models. 
They describe the propagation of particles in a curved background, just like 
2D nonlinear sigma models describe the propagation of strings\footnote{
The requirement of conformal invariance restricts the possible backgrounds 
on which the string propagates, but no such  requirement is present for 
the particle case, at least for spin 0 and 1/2.}.
Sigma models are super-renormalizable in 1D, and their UV structure
together with the necessary renormalization conditions (which produce
explicit counterterms) have been extensively discussed in the literature using 
various regularization methods 
\cite{Bastianelli:1998jm,DeBoer:1995hv,Bastianelli:2000nm}.

In this paper we plan to address in a more systematic way the infrared issue 
related to the treatment of zero modes appearing on the 
circle and the interplay with the reparametrization invariance of 
nonlinear sigma models.

Different ways of treating such zero modes have been developed in the 
literature. For example in \cite{Fliegner:1997rk}
an arbitrary background charge function $\rho(\tau)$ was used
to interpolate between various boundary conditions, thus lifting the zero
modes in different ways.
There it was shown that for models in flat space the effective lagrangian
calculated with different background charges $\rho$ differed only by 
total derivatives, thus producing the same effective action.
While this is not causing any particular problem in flat space
(total derivatives are even beneficial in certain cases, since they allow
to cast the effective action in a more compact form), 
the naive extension of this method to curved space 
was seen to introduce noncovariant total derivatives  
\cite{Schalm:1998ix,Bastianelli:2002fv}, raising questions 
about the correct use of Riemann normal coordinates to simplify 
calculations.

A general method for dealing with these zero modes has been developed 
by Friedan in his treatment of 2D bosonic sigma models \cite{Friedan:1980jm}, 
and recently employed in the 1D case by Kleinert and Chervyakov 
\cite{Kleinert:2003zq}, where the comparison between  the so-called 
SI (string inspired) propagator and the DBC (Dirichlet boundary conditions) 
propagator was carried out to show how the former could produce a 
covariant result as the latter. The string inspired 
propagator \cite{Strassler:1992zr} is translational invariant on the worldline,
while the DBC one is not.

Here we review and analyze the treatment of
the  zero modes to clarify it further and resolve some remaining puzzles.
We use BRST methods to factor out the zero modes.
The BRST symmetry is related to the gauge fixing of a shift 
symmetry. It is the shift symmetry typical of background field methods
\cite{Abbott:1980hw}.
However, in the present case the ``background field'' is integrated over 
in the path integral (i.e. it is made dynamical), and thus the shift 
symmetry must be gauge fixed.
An immediate result of this procedure of extracting the zero modes is that
different gauges are guaranteed to produce the same effective 
action, implying that effective lagrangians can only differ by 
total derivatives, even in curved space.
However, the total derivatives will in general be noncovariant.
On the other hand, suitable covariant gauges will naturally produce effective 
lagrangians differing from each other only by covariant total derivatives.
Various examples are extensively analyzed to test these predictions.
In particular, we show how the calculation of the trace of the heat kernel 
and the related Seeley--DeWitt coefficients, which identify the effective 
action for a scalar particle in the proper time expansion \cite{DW}, 
is achieved with different background charges. 

We also extend the zero mode treatment to the supersymmetric case.
This is relevant for the worldline description of spin 1/2 fermions
coupled to gravity \cite{Bastianelli:2002qw}.
We use it for an explicit check that the chiral anomaly does not receive 
higher order worldline contributions.
In fact, one may recall that 
the $N=1$ supersymmetric nonlinear sigma model  was used in 
\cite{Alvarez-Gaume:1983at} to compute the chiral anomaly of a spin 1/2 
field. The computation was based on the fact that the chiral 
anomaly could be identified as the Witten index of the corresponding 
supersymmetric quantum mechanical model \cite{Witten:df}.
Supersymmetry  implies that higher order worldline contributions should not 
modify the value of the Witten index, and we test this  explicitly
using 1D path integrals.

The paper is organized as follows. In section 2 we discuss bosonic 
nonlinear sigma models. As a test we calculate perturbatively the trace of 
the heat kernel and the related Seeley--DeWitt coefficients
with different background charges.
In section 3 we consider supersymmetric nonlinear sigma models
and use them to study worldline corrections to the susy 
quantum mechanical computation 
of the chiral anomaly. In section 4 we present our conclusions.
For completeness and further clarifications we present the
simpler case of a bosonic linear sigma model in 
appendix~\ref{sub:linear-sigma}.
Other appendices include conventions and integrals needed for the computations
described in the text.

\section{Bosonic nonlinear sigma models}
\label{section:nonlinear-sigma}

Let us consider the partition function of the 1D nonlinear sigma model
\bea
Z(\beta) = \oint {\cal D}x\  e^{-S[x]}  \ ,  \quad \quad S[x]={1\over \beta}
\int_0^1 \!\!d\tau \left [
{1\over 2}g_{\mu\nu}(x)\dot x^\mu \dot x^\nu +\beta^2 V(x) \right ] 
\label{1.act}
\eea
where $g_{\mu\nu}$ and $V$ are a metric and a scalar potential
defined on target space, which we take to be $D$  dimensional.
The path integral is computed with periodic boundary conditions (PBC),
i.e. on the circle. We use euclidean time, and 
it is well-known that periodic boundary conditions then 
yield the statistical partition function.
The circle is just the loop made by the particle in target space. 
It can be parametrized by $t\in [0,\beta]$, with $\beta$
the total length of the circle. 
We also use a rescaled proper time $\tau=t/\beta$ and this
rescaling explains the factor ${1\over \beta}$ multiplying the action
as well as the $\beta^2$ factor in front of the scalar potential.

The partition function $Z(\beta)$ is sometimes called the trace 
of the heat kernel\footnote{The operator
$e^{-\beta H}$ is called the heat kernel, and
the partition function is given by 
$Z(\beta) = \Tr e^{-\beta H}= \oint {\cal D}x\  e^{-S[x]} $, 
where $S[x] $ is the action corresponding to 
the model with quantum hamiltonian $H$.} \cite{DW}.
It is related to the one-loop effective action of a scalar field 
with kinetic operator  $ -\Box +2V + m^2$ ( 
$\Box$ is the covariant scalar laplacian depending on the metric
$g_{\mu\nu}$  and $m$ is the mass of the scalar particle)
 by an integral over the proper time $\beta$
\bea
\Gamma[g,V] = -{1\over 2} \int_0^\infty {d\beta\over \beta } 
\, e^{-{1\over 2}m^2 \beta} \,
Z(\beta)  \ .
\label{ea}
\eea

In a perturbative computation zero modes appear. 
In fact for $g_{\mu\nu}=\delta_{\mu\nu}$ and $V=0$ the action is
invariant under the constant translations 
$\delta x^\mu (\tau) = \epsilon^\mu $. Hence the volume of target space
appears as a factor, just like the volume of a gauge group.
Background fields generically break this translation invariance. 
Nevertheless it is both useful and necessary to extract from the path integral
the ``collective coordinates''  or ``center of mass'' of the loops 
$x^\mu (\tau)$, which for simplicity we continue to call zero modes.
It is useful, since it allows to produce the partition function as
an integral of a partition function density.
It is necessary, since in perturbative calculations around the free action 
one needs to invert the free kinetic term to obtain the perturbative 
propagator.
A general method for treating the zero modes for nonlinear sigma models 
has been developed by Friedan \cite{Friedan:1980jm}, and employed recently 
in the 1D case by Kleinert and Chervyakov \cite{Kleinert:2003zq}.
Useful references are also \cite{Fradkin:1984pq,Howe:vm}.

Let us rederive these results and extend them by using an arbitrary 
background charge $\rho(\tau)$.
To extract the zero modes one can proceed as 
follows\footnote{In appendix~\ref{sub:linear-sigma} we describe the simpler 
case of a linear sigma model using both Faddeev--Popov and BRST methods.}.
The action of the nonlinear sigma model $S[x(\tau)]$ depends on the periodic 
paths $x^\mu(\tau)$ which describe loops with the topology of 
a circle in target space.
One may introduce a redundant variable $x_0^\mu$
by setting $x^\mu(\tau)= x^\mu_0+y^\mu(\tau)$ 
in the action, $S[x(\tau)]= S[x_0+y(\tau)]$. 
Of course this automatically introduces the shift symmetry
\bea
\delta x^\mu_0 &=& \epsilon^\mu 
\nonumber\\
\delta y^\mu(\tau) &=& -\epsilon^\mu 
\label{linsplit}
\eea
as in the background field method \cite{Abbott:1980hw}.
We only consider constant $x^\mu_0$ so that 
the shift symmetry requires a constant parameter $\epsilon^\mu$.
However, contrary to the background field method,
we now consider both $x_0^\mu$ and $y^\mu(\tau)$ 
as dynamical variables (i.e. to be integrated over in the path-integral).
The shift symmetry is thus promoted to a gauge symmetry and must be gauge
fixed since each physical configuration has to be counted only once.
To gauge fix we use BRST methods and introduce a constant ghost field 
$\eta^\mu$ together with the following BRST transformation rules
\bea
\delta x^\mu_0 &=& \eta^\mu \Lambda 
\nonumber\\
\delta y^\mu (\tau) &=& -\eta^\mu  \Lambda 
\nonumber\\
\delta \eta^\mu &=& 0 \ .
\eea
To fix a gauge one must also introduce constant nonminimal 
fields $\bar \eta_\mu, \pi_\mu$ with the BRST rules
\bea
\delta \bar \eta_\mu = i \pi_\mu \Lambda \ ,\quad\quad\quad
\delta \pi_\mu = 0  \ .
\eea
Choosing the gauge fixing fermion 
\be
\Psi = \bar \eta_\mu \int_0^1 \!\!d\tau \,\rho(\tau ) y^\mu(\tau )
\label{c1.gferm}
\ee
which depends on the arbitrary function $\rho(\tau )$
normalized to $\int_0^1 d\tau \rho(\tau )=1$
produces the following gauge fixed action
\bea 
S_{gf} [x_0, y,\eta,\bar\eta,\pi]
&=& S[x_0 +y] + {\delta\over \delta \Lambda}\Psi 
\nonumber\\
&=& S[x_0 +y] +  i \pi_\mu \int_0^1 \!\! d\tau \,\rho(\tau ) y^\mu(\tau )
-\bar \eta_\mu \eta^\mu
\eea
where ${\delta\over \delta \Lambda}$ denotes a BRST variation
with the anticommuting parameter $\Lambda$ removed form the left. 
In this gauge the ghosts can be trivially integrated out, 
while the integration over the auxiliary variable $\pi_\mu$ produces 
a delta function which constrains the fields $y^\mu$ to satisfy 
\be
\int_0^1 \!\! d\tau\, \rho(\tau) y^\mu(\tau)=0 \ .
\label{constraint}
\ee 
With this constraint the perturbative kinetic term for the periodic 
fields $y^\mu(\tau)$, proportional to ${d^2\over d\tau^2}$, can be 
inverted to obtain the propagator. The BRST symmetry implies 
that the partition function is independent of the gauge 
parameter $\rho$. 
The specific case of $\rho(\tau)=\delta(\tau)$ gives the DBC
propagator since $y^\mu(0)=y^\mu(1)=0$. The case $\rho(\tau)=1$ gives instead 
the SI propagator since
now the center of mass is absent from the fluctuations $y^\mu$,
see e.g. the discussion in  \cite{Fliegner:1997rk}.

Thus the partition function is independent of $\rho$ and 
can be expressed as an integral over the zero modes
\bea
Z(\beta)  = \int d^Dx_0 \, {\sqrt{g (x_0)} \over (2 \pi \beta)^{D\over 2}}\, 
z^{(\rho)}(x_0,\beta) 
\label{1pf1}
\eea
where the factor ${\sqrt{g (x_0)} \over (2 \pi \beta)^{D\over 2}} $
has been extracted  for convenience from the definition of
$z^{(\rho)}(x_0,\beta)$.
Note however that the density $z^{(\rho)}(x_0,\beta)$  may in general 
depend on $\rho$. This can only happen through total derivatives which
must then integrate to zero.

Although this is correct, at least formally, there are some practical problems.
The constraint arising
from the gauge fixing does not have simple transformations rules under 
change of coordinates (coordinate differences like 
$y^\mu = x^\mu-x_0^\mu$ do not transform as vectors).
This causes some technical problems when one wants to check the explicit 
covariance of the final  result. In particular, one explicitly
finds in $z^{(\rho)}(x_0,\beta)$
total derivatives which depend on the choice of $\rho$ 
and are {\em not covariant} under change of the coordinates $x^\mu_0$
\cite{Schalm:1998ix,Bastianelli:2002fv}.  
Let us recall that DBC are related to the calculation of the heat kernel
even for non coinciding points, and they are known to give a covariant
result. The other propagators related to different background charges 
produce instead noncovariant  total derivatives.
A calculation at order $\beta$ using both the DBC and SI propagators 
was presented in \cite{Bastianelli:2002fv} to explicitly identify
the noncovariant total derivative term appearing at that order.
For general $\rho$ the expression given in \cite{Bastianelli:2002fv} 
generalizes to 
\be
z^{(\rho)}(x_0,\beta)
= 1 +\beta\, \Big  ({ 1\over 12 } R - V \Big) 
+ {\beta \over 2 \sqrt{g}} 
\Big (C_\rho + {1\over 12}\Big )
\partial_\mu ( \sqrt{g} g^{\alpha\beta}
\Gamma_{\alpha\beta}{}^\mu )  + O(\beta^2)  
\label{gntd}
\ee
where the precise value of $C_\rho$ is defined later in eq. (\ref{defd}).
The appearance of noncovariant total derivatives 
may raise doubts about the use of Riemann normal coordinates
which are often used to simplify calculations. In fact, in  
\cite{Schalm:1998ix} the assumption of a naive use of Riemann normal 
coordinates was seen to produce the wrong trace anomaly in 2 and 4 
dimensions.

Riemann normal coordinates can nevertheless be used, as showed by Friedan 
in his discussion of nonlinear sigma models \cite{Friedan:1980jm}.
Friedan noticed that the simple linear shift symmetry (\ref{linsplit})
becomes nonlinear when using Riemann normal coordinates $\xi^\mu$ 
centered at $x_0^\mu$.
One should use this nonlinear shift symmetry to correctly 
perform the gauge fixing in Riemann coordinates.
Riemann normal coordinates 
have the property that they are manifestly covariant under 
reparametrization of the point $x_0^\mu$.
Now the action $S[x_0,\xi(\tau)] \equiv S[x_0+y(x_0,\xi(\tau))]$ 
is invariant under the nonlinear shift symmetry
\bea
\delta x^\mu_0 &=& \epsilon^\mu 
\nonumber\\
\delta \xi^\mu(\tau) &=& - Q^\mu{}_\nu(x_0,\xi(\tau))\epsilon^\nu  
\label{nonlinear}
\eea
which is a reformulation of (\ref{linsplit}) in these new coordinates.
Note that since the origin of the Riemann normal coordinates is shifted, 
the nonlinear transformation is defined in such a way that the
 new fields ${\xi^\mu}' = \xi^\mu +\delta \xi^\mu$ 
are expressed in Riemann normal coordinates defined around 
the new origin $x^\mu_0{}' = x^\mu_0 + \delta x^\mu_0$.
The expression $Q^\mu{}_\nu(x_0,\xi)$ can be explicitly calculated 
and can be found in \cite{Friedan:1980jm}.
We report it here up to the order needed in subsequent calculations
\bea
Q^\mu{}_\nu(x_0,\xi)&=&
\delta^\mu_\nu + 
{1\over 3} R^\mu{}_{\alpha\beta\nu}\, \xi^\alpha \xi^\beta 
+ {1\over 12} \nabla_\gamma R^\mu{}_{\a\b\nu} \, \xi^\alpha \xi^\b 
\xi^\gamma \nonumber\\
&&  + \Big({1\over 60} 
\nabla_\gamma \nabla_\delta R^\mu{}_{\alpha\beta\nu} 
-{1\over 45} R^\mu{}_{\alpha\beta\lambda} R^\lambda{}_{\gamma\delta\nu}
\Big)\xi^\alpha \xi^\beta \xi^\gamma \xi^\delta   + O(\xi^5) \ .
\label{one}
\eea

Now we can introduce ghosts and auxiliary fields as usual. 
The BRST symmetry for the nonlinear shift symmetry is 
\bea
\delta x^\mu_0 &=& \eta^\mu \Lambda 
\nonumber\\
\delta \xi^\mu (\tau) &=& - Q^\mu{}_\nu(x_0,\xi(\tau))\, 
\eta^\nu \Lambda 
\nonumber\\
\delta \eta^\mu &=& 0 
\nonumber\\
\delta \bar \eta_\mu &=& i \pi_\mu \Lambda 
\nonumber\\
\delta \pi_\mu &=& 0~.
\eea
It is nilpotent since $Q^\mu{}_\nu(x_0,\xi(\tau))$ satisfies certain 
relations arising from the abelian nature of the shift symmetry.
Using the gauge fermion
\be
\Psi = \bar \eta_\mu \int_0^1 \!\! d\tau \,\rho(\tau ) \xi^\mu(\tau )
\ee
produces the gauge fixed action
\bea 
S_{gf} [x_0,\xi,\eta,\bar\eta,\pi]
&=& S[x_0,\xi] + {\delta\over \delta \Lambda}\Psi 
\nonumber\\
&=& S[x_0,\xi] +  i \pi_\mu \int_0^1 \!\! d\tau \,\rho(\tau ) \xi^\mu (\tau )
\nonumber\\
&&-\bar\eta_\mu \int_0^1 \!\! d\tau \,\rho(\tau ) 
Q^\mu{}_\nu(x_0, \xi(\tau ))\, 
\eta^\nu  \ .
\label{gf2}
\eea
The integration over the auxiliary variable $\pi_\mu$ gives again a 
delta function which constrains the fields $\xi^\mu$ to satisfy 
\be
\int_0^1 \!\! d\tau\, \rho(\tau) \xi^\mu(\tau)=0 
\label{con2}
\ee 
so that their propagator can be obtained.
This constraint has a simple tensorial transformation law under the change 
of coordinates of $x_0^\mu$, in fact these coordinates transforms 
as vectors under a reparametrization of the origin $x_0^\mu$.
The ghosts now give a nontrivial contribution, i.e. a nontrivial 
Faddeev--Popov determinant.

The above gauge fixed actions can be used in the path integral.
Of course, one also needs to use a path integral measure that is both 
reparametrization and BRST invariant. This is given by
\bea
{\cal D}x &=&
\prod_{0\leq\tau<1} \sqrt{g(x(\tau))} dx(\tau) \ 
\longrightarrow \ 
dx_0 \, d\eta \, d\bar \eta \, d\pi 
\prod_{0\leq\tau<1} \sqrt{g(x(\tau))} dx(\tau) 
\nonumber\\
&=& 
dx_0 \, d\eta \, d\bar \eta \, d\pi 
\prod_{0\leq\tau<1} \sqrt{g(x_0+ y(\tau))} dy(\tau)  \ .
\eea
where we have first added to the sigma model measure
the measure for the BRST quartet 
$ x_0^\mu, \eta^\mu, \bar \eta_\mu , \pi_\mu$ which is formally 
identical to unity
(two commuting fields give a volume which is compensated by that of 
the two anticommuting fields), and then performed the
change of variables identified by the ``background-quantum'' split
$x^\mu=x_0^\mu+y^\mu$.
We have here used the linear splitting, but we could equally well
use  the nonlinear one written in terms of the Riemann normal coordinates
centered at $x_0^\mu$.
 In fact the measure is reparametrization invariant,
i.e. of the same form in any coordinate system
\bea
{\cal D}x &=& dx_0 \, d\eta \, d\bar \eta \, d\pi 
\prod_{0\leq\tau<1} \sqrt{g(x_0, \xi(\tau))} d\xi(\tau)~.
\eea
Note that with $g(x_0, \xi(\tau))$ we have indicated the determinant of the
metric in Riemann normal coordinates centered at $x_0$.
For future reference we list the expansion of the metric
in Riemann coordinates centered at $x_0$
up to the order needed in later calculations
 \bea
g_{\mu\nu} (x_0, \xi)  &=& 
g_{\mu\nu} (x_0) + {1\over 3} R_{\mu\alpha\beta\nu}(x_0) \xi^\a \xi^\b
+{1\over 6} \nabla_\gamma R_{\mu\alpha\beta\nu}(x_0) 
\xi^\alpha \xi^\beta \xi^\gamma
\nonumber\\
&& + \Big ( {1\over 20} 
\nabla_\delta \nabla_\gamma R_{\mu\alpha\beta\nu}(x_0) 
+ {2\over 45}  R_{\mu\alpha\beta}{}^{\s} R_{\s\gamma\delta\nu}(x_0) 
\Big ) \xi^\alpha \xi^\beta \xi^\gamma \xi^\delta
+ O(\xi^5)  \ .\quad\quad  
\label{two}
  \eea

It is now useful to introduce additional ghosts to exponentiate
the nontrivial part of the measure.
We use commuting $a^\mu$ and anticommuting 
$b^\mu, c^\mu$ ghost fields to reproduce the correct measure
\cite{Bastianelli:1991be}
\bea
\prod_{0\leq\tau<1} \sqrt{g(x_0, \xi(\tau))} d\xi(\tau)
&=& \prod_{0\leq\tau<1} d\xi(\tau)
\int  \prod_{0\leq\tau<1} d a(\tau) d b(\tau) d c(\tau)
\, e^{-S_{msr}}\\
 S_{msr}[\xi,a,b,c] 
&=& {1\over \beta} \int_0^1 \!\! d\tau \left [
{1\over 2}g_{\mu\nu}(x_0, \xi)(a^\mu a^\nu + b^\mu c^\nu) \right ]  \ .
\eea
The extra vertices arising from the measure will contribute together with
similar vertices from the sigma model action to make the final result finite
\cite{Lee:vm}.

We are now ready to re-assemble all parts of the path integral
with the zero modes factored out by the nonlinear shift symmetry
\bea
Z(\beta)= \int dx_0 \, d\eta \, d\bar \eta \, d\pi 
\oint D\xi Da Db Dc \,  e^{- S_{gf} [x_0,\xi,\eta,\bar\eta,\pi] 
- S_{msr}[x_0,\xi,a,b,c] } \ .
\eea
The auxiliary field $\pi^\mu$ can be integrated out to obtain
\bea
Z(\beta)= \int dx_0 \, d\eta \, d\bar \eta  
\oint D\xi Da Db Dc \ \delta \Big( 
\int^1_0 \!\! d\tau\, \rho(\tau) \xi^\mu(\tau)\Big)\, e^{- S_q}
\eea
where 
\be
S_q= S_{gf} [x_0,\xi,\eta,\bar\eta] + S_{msr}[x_0,\xi,a,b,c] 
\ee
with the auxiliary field $\pi_\mu$ eliminated from $S_{gf}$.
Finally, when perturbation
around the leading terms in (\ref{one}) and (\ref{two}) 
is appropriate, one immediately obtains the following perturbative expansion
\bea
Z(\beta)
= \int d^Dx_0 {\sqrt{g (x_0)} \over (2 \pi \beta)^{D\over 2}}\  
\la \exp (- S_q^{(int)}) \ra
\label{1.last}
\eea
where the expectation value of the 
interactions are computed with the propagators
\bea
\la\xi^\mu(\tau)\xi^\nu(\sigma)\ra
&=& 
-\beta g^{\mu\nu}(x_0) \B_{(\rho)}(\tau,\sigma) \nonumber\\[.5mm] 
\la a^\mu(\tau) a^\nu(\sigma)\ra 
&=&  
\beta g^{\mu\nu}(x_0) \Delta_{gh} (\tau-\sigma) \nonumber\\[.5mm] 
\la b^\mu(\tau) c^\nu(\sigma)\ra 
&=& 
-2\beta g^{\mu\nu}(x_0) \Delta_{gh} (\tau-\sigma) \nonumber\\[.5mm] 
\la \eta^\mu \bar \eta_\nu \ra 
&=& 
- \delta^\mu_\nu      \ .
\label{generic_prop}
\eea
The terms in the measure can be traced back as follows:
the factor $\sqrt{g (x_0)}$ is due to the $a,b,c$ ghosts which contain
the constant ``zero modes'', the factor 
$(2 \pi \beta)^{-{D\over 2}}$ is the usual free particle measure
which corresponds to the determinant of $-{1\over 2 \beta}
 {d^2\over d\tau^2} $ on the circle with zero modes excluded.
The Green functions appearing in the propagators are as follows.
${\cal B}_{(\rho)}(\tau,\sigma)$ is the Green function
of the operator $ {d^2\over d\tau^2}$  acting on fields
constrained by the equation $ \int_0^1 d\tau \rho(\tau) \xi^\mu(\tau)=0$.
It depends on $\rho$ and satisfies
\bea
{d^2\over d\tau^2}{\cal B}_{(\rho)}(\tau,\sigma) = \delta(\tau-\sigma) -
\rho(\tau) \ .
\eea
It is explicitly given by \cite{Fliegner:1997rk}
\bea
{\cal B}_{(\rho)}(\tau,\sigma) 
= \Delta(\tau-\sigma) - F_\rho(\tau)
-F_\rho(\sigma) + C_\rho 
\label{green}
\eea
where
\bea
&&\Delta(\tau -\sigma) = {1\over 2} |\tau-\sigma| -
{1\over 2} (\tau-\sigma)^2 -{1 \over 12}
\nonumber\\[1mm]
&&F_\rho(\tau) = \int_0^1 \!\! dx\ \Delta(\tau-x)\rho(x) \ ,\quad
C_\rho = \idt F_\rho(\tau) \rho(\tau) \ ,
\label{defd}
\eea
and it clearly satisfies 
\bea
\idt \rho(\tau)\, {\cal B}_{(\rho)}(\t,\s)=0 \ .
\eea
Note that the auxiliary function $\Delta(\tau -\sigma)$ is the 
unique translational invariant Green function on the circle
(the ``string inspired'' propagator
of \cite{Strassler:1992zr}, which corresponds to $\rho(\tau)=1$). 
In the following we will simply write ${\cal B}_{(\rho)} = \cal B$ 
as no confusion can arise.
The Green function for the ghosts is given by the delta function
\be 
\Delta_{gh} (\tau-\sigma) = \delta(\tau-\sigma) 
\ee
but we continue to call it $\Delta_{gh}$ as in perturbative
calculations it appears in a regulated form.

We are now ready to test this set up. To summarize, 
the expectations are that the partition function $Z(\beta)$
will not depend on $\rho$, but that the density 
$z^{(\rho)}(x_0,\beta) $  will in general be $\rho$-dependent 
through total derivatives which integrate to zero. 
Moreover, using 
Riemann normal coordinates and the associated 
nonlinear shift invariance
to extract the integral over the zero modes $x^\mu_0$,
one expects $z^{(\rho)}(x_0,\beta) $ 
to be covariant under change of coordinates of $x^\mu_0$, 
so that the specific total derivatives which may eventually appear will 
also  be covariant.

In the next subsections we explicitly verify, using Riemann normal
coordinates, that these total derivatives term are non-zero and
covariant also in the presence of external potentials like $V$.

\subsection{Partition function at 3 loops}

We present here the explicit perturbative calculation of the partition
function density to order $\b^2$ using Riemann normal coordinates (RNC)
and dimensional regularization on the worldline. 
The nonlinear sigma model in one dimension  is super-renormalizable 
and one needs to choose a specific regularization
scheme to compute unambiguously the perturbative expansion.
We use dimensional regularization which 
requires an explicit counterterm $V_{DR}=-\frac{1}{8}R $ to 
guarantee that the sigma model in (\ref{1.act})
will have $H=-{1\over 2} \Box + V$ as quantum hamiltonian
\cite{Bastianelli:2000nm}. 
In the following we will use the rules for dimensional regularization
explained in \cite{Bastianelli:2000nm}.
Dimensional regularization has also been discussed for 1D nonlinear
sigma model with infinite proper time in \cite{Kleinert:1999aq}.

The partition function density in eq. (\ref{1pf1}) and
(\ref{1.last}) can be expressed in terms of connected worldline graphs as
\be
z^{(\r)}(x_0,\b)  =  \la \exp(-S_q^{(int)})\ra  = 
   \exp (\la e^{- S_q^{(int)} }\ra_c -1 )
\label{2.1exp}
\ee
where $S_q \equiv S_{gf}+S_{msr}$ is the full quantum action,
$\la ... \ra_c $ denotes connected graphs, and 
the propagators are given in eq. (\ref{generic_prop}).

In order to appreciate the contribution from the FP determinant, 
we separate the corresponding action
\bea
S_{FP} = - \bar\eta_\mu   \idt \r(\t) 
   Q^\mu_\nu(x_0,\xi(\t)) \, \eta^\nu
\label{SFP}   
\eea
from the other contributions, gathered in $\bar S \equiv S_q-S_{FP}$. 
As usual, we organize the interaction terms in such a way that 
$S_q^{(int)}= S_q-S_{q,2} = S_{q,4}+ S_{q,5}+\ldots$, where
$S_{q,n}=\bar S_n+S_{FP,n}$ gives rise to vevs of order 
$\b^{\frac{n}{2}-1}$ (recall that $S_{q,3}=0$
in RNC, see eqs. (\ref{one}) and (\ref{two}) ). 
Hence
\be
\la e^{- S_q^{(int)} }\ra_c -1= \Big \la -S_{q,4} -S_{q,6}
+\frac{1}{2}S_{q,4}^2 \Big \ra_c +O(\b^3) \ .
\label{expansion}
\ee 
We denote by $V_q $ the potential which includes the counterterm 
arising in dimensional regularization
\be 
V_q= V+ V_{DR}=V -{1\over 8}R \ .
\ee

We now list the results for the various terms
appearing in (\ref{expansion})
and report in the appendix~\ref{sub:integrals} the expressions and 
values of the relevant connected one- and two-loop worldline integrals 
$H_i$, calculated in dimensional regularization
\bea
-\la \bar S_4 \ra &=& -\frac{\b}{6}(-H_1+H_2) R - \b V_q   =
    \b \left\{ \left(\frac{C_\r}{3}-\frac{1}{72}\right)R - V_q      
    \right\}     
\nonumber\\ [2mm]
-\la S_{FP,4} \ra &=& \frac{\b}{3}H_3 R  =
    -\b\left(\frac{C_\r}{3}+\frac{1}{36}\right)R           
\nonumber\\ [2mm]
-\la \bar S_6 \ra &=& \b^2 \left\{ \frac{1}{20} (-H_4+H_5)\ \Box R
    +\frac{1}{45} (H_4-H_5) \left [ R^2_{\mu\nu}+ \frac{3}{2}
     R^2_{\mu\nu\a\b}\right ] +\frac{1}{2}H_1\ \Box V_q  \right\}  
\nonumber\\ [.5mm]
  &=& \b^2 \left\{ \left(2C_\r^2 -\frac{7}{18}C_\r -D_\r 
    -\frac{5}{6}E_\r +\frac{1}{144}\right)
    \left [ -\frac{1}{20}\Box R +\frac{1}{45}R^2_{\mu\nu}\right.\right.
\nonumber\\[.5mm]
  &&\left.\left.  +\frac{1}{30} R^2_{\mu\nu\a\b} \right ]  
     +\frac{1}{2} \left(C_\r-\frac{1}{12}\right) \Box V_q  \right\} 
\nonumber\\ [2mm]
\frac{1}{2}\la \bar S_4^2 \ra_c  &=& \b^2 \left\{ \frac{1}{36}(H_6+H_7
    +2H_8+2H_9-4H_{10}-4H_{11}+2H_{12}) R^2_{\mu\nu} \right.
\nonumber\\ [.5mm]
  && \left. 
    +\frac{1}{24}(H_{13}+H_{14}-2H_{15}) R^2_{\mu\nu\a\b}  \right\}     
\nonumber\\ [.5mm]
  &=& \b^2 \left\{ \left(\frac{1}{18}C_\r^2 +\frac{5}{162}C_\r 
    +\frac{2}{27}E_\r -\frac{7}{12960} \right) R^2_{\mu\nu} \right. 
\nonumber\\ [.5mm]
  && \left. +\left(-\frac{1}{12}C_\r^2 +\frac{1}{54}C_\r
    +\frac{1}{9}E_\r +\frac{13}{8640} \right) R^2_{\mu\nu\a\b} \right\}
\nonumber\\ [2mm]
-\la S_{FP,6} \ra &=& -\b^2 \left\{\frac{1}{30}H_{16}\ \Box R 
    +\left(\frac{1}{45}H_{16}+\frac{1}{18}H_3^2\right)R^2_{\mu\nu}
    \right.\nonumber\\[.5mm]
    &&\left. +\left(\frac{1}{30}H_{16}+\frac{1}{12}H_{17}\right)
    R^2_{\mu\nu\a\b}\right\}   
\nonumber\\ [.5mm]
  &=& \b^2 \left\{ \left [\frac{1}{10}C_\r^2 -\frac{1}{180}C_\r  
    -\frac{2}{15}D_\r -\frac{1}{4320} \right] \Box R \right.
\nonumber\\ [.5mm]    
&&  + \left [ \frac{1}{90}C_\r^2 -\frac{7}{540}C_\r 
    -\frac{4}{45}D_\r-\frac{7}{12960} \right] R^2_{\mu\nu}  
\nonumber\\ [.5mm]
  && \left. +\left [\frac{1}{60}C_\r^2-\frac{1}{180}C_\r 
    +\frac{1}{30}D_\r -\frac{1}{12}E_\r -\frac{1}{2880}
     \right] R^2_{\mu\nu\a\b} \right\}
\nonumber\\ [2mm]
\frac{1}{2}\la S_{FP,4}^2 \ra_c &=& \frac{\b^2}{9}H_{17}\ R^2_{\mu\nu}
   = \frac{\b^2}{9} \left(C_\r^2 -2D_\r +E_\r +\frac{1}{720}\right)
     R^2_{\mu\nu}                                              
\nonumber\\ [2mm]
\la S_{FP,4}\, \bar S_4 \ra_c &=& \frac{\b^2}{9} (H_{18}+H_{19}-2H_{20}) \ 
   R^2_{\mu\nu}                                                 
\nonumber\\ [.5mm]
&=&  
\frac{\b^2}{9} \left(-2C_\r^2-\frac{1}{12}C_\r 
    +3D_\r -\frac{3}{2}E_\r -\frac{1}{180}\right) R^2_{\mu\nu}  \ .
\label{2.1val}
\eea
All tensors appearing here are evaluated at $x_0$.

As a check on these results, 
note that all contributions from the FP determinant 
vanish when using DBC.
In fact, setting  $\rho(\tau)=\d(\tau)$ one obtains for the
$\rho$ dependent coefficients $C_\rho, D_\rho, E_\rho$ 
the following values $C_{DBC}= -{1\over 12}, 
D_{DBC}= {1\over 144}, E_{DBC}=-{1\over 180} $. 
This is expected, since for $\rho(\tau)=\d(\tau)$ 
the constraint (\ref{con2}) enforces 
the DBC, i.e.  $ \xi^\mu(0)= \xi^\mu(1)=0 $.
This in turn implies that the  FP determinant in eq. (\ref{gf2}) 
becomes trivial,  since 
$Q^\mu{}_\nu(x_0,0)= \delta^\mu_\nu $, as seen form  eq. (\ref{one}). 
Note also that we have defined the $\rho$-dependent coefficients to 
vanish in the SI case, i.e. when $\rho(\tau)=1$.

We can now insert the results (\ref{2.1val}) into 
(\ref{expansion}) and (\ref{2.1exp}) to obtain
the partition function density valid to the order $\b^2$ 
\bea
z^{(\rho)}(x_0,\b) &=&  \exp\left\{-\b\left [{1\over 24}R + V_q  \right ]
   +\b^2 \left [\left( -\frac{1}{144}C_\r -\frac{1}{12}D_\r 
   +\frac{1}{24}E_\rho +\frac{1}{864}\right) \Box R  \right.\right. 
\nonumber\\[.5mm]
  && \left.  + \left(C_\rho-\frac{1}{12}\right)
   \left({1\over 48}\Box R+ {1\over 2}\Box V_q   \right)
  -\frac{1}{720} R^2_{\mu\nu} +\frac{1}{720} R^2_{\mu\nu\a\b} \right ] 
\nonumber\\[.5mm]  
&&\left. + O(\beta^3)\right\}~. 
\label{zrho} 
\eea
Expanding the exponent to order $\b^2$, one sees that only covariant 
total derivatives have $\rho$-dependent coefficients
\bea
z^{(\rho)}(x_0,\b) &=& 1 -\b \left [{1\over 24}R + V_q  \right ]
   +\b^2 \left [ {1\over 2} \left ({1\over 24}R + V_q  \right )^2
+\frac{1}{720} R^2_{\mu\nu\a\b}   -\frac{1}{720} R^2_{\mu\nu} 
\right. 
 \nonumber\\ [.5mm]
  && \left.  +\left( \frac{1}{72}C_\r -\frac{1}{12}D_\r 
   +\frac{1}{24}E_\rho -\frac{1}{1728}\right) \Box R  
+ {1\over 2} \left(C_\rho-\frac{1}{12}\right) \Box V_q   \right ] 
\nonumber\\ 
&&+ O(\beta^3)~. 
\label{zrho1} 
\eea

Let us discuss some consequences of this formula.
We can integrate this density over the zero modes $x_0$
to obtain the partition function
\be
Z(\beta) =
\int d^Dx_0 \, {\sqrt{g (x_0)} \over (2 \pi \beta)^{D\over 2}}\, \Big(
1+ a_1(x_0)\beta +  a_2(x_0)\beta^2 + ... \Big ) 
\ee
where $a_n(x_0)$ are the so-called Seeley--DeWitt coefficients.
Inserted into (\ref{ea}) this gives in turn 
the effective action of the scalar field with kinetic term
$ -\Box +2V + m^2$. 
It is well-known that this proper time expansion of the effective action
fails for massless fields. In fact for vanishing mass the damping factor
$ e^{-{1\over 2}m^2 \beta}$, which guarantees convergence in eq. (\ref{ea}),
becomes unity. 
Nevertheless the Seeley--DeWitt coefficients are still useful in this
case as well, as they give the counterterms needed to renormalize the 
one-loop effective action. Moreover, for conformal fields in $D$
dimensions the coefficient $a_{D\over 2}$ gives the local
trace anomaly \cite{DW}.

The main points to stress are: 

$\bullet$ 
For the validity of the perturbative calculation of $Z(\beta)$ 
we have to assume that all external fields describing the interactions
should vanish sufficiently fast at infinity.
Thus all total derivatives integrate to zero. Therefore gauge independence
is verified.

$\bullet$ 
The covariant local expansions of $z^{(\rho)}(x_0,\b)$ 
in Riemann normal coordinates (RNC) are different 
for different $\rho$'s. The difference is given by total 
derivatives with  coefficients depending on $\rho$ and which are explicitly 
nonvanishing. From the computation to order $\beta $  
in \cite{Kleinert:2003zq} this could not be evinced. In fact at that 
order there is no covariant total derivative 
with the correct dimensions that could possibly contribute.
Thus, at order $\beta$  different $\rho$'s
produce the same local expansion.
In principle there could have existed  hidden relations guaranteeing
the same local expansion of $z(x_0,\b)$ for different $\rho$'s
at any order in $\beta$.
We see explicitly that this is not the case and covariant total derivative 
may arise.
  
$\bullet$
One can use these results to compute trace anomalies, which in $D$ dimensions 
are given by the Seeley--DeWitt
coefficient $a_{D\over 2}$ associated to the corresponding 
conformal operator.  
In $D=2$ there cannot be any covariant total derivative contribution 
to $a_1$, so the local trace anomaly can be obtained with any $\rho$.
In $D=4$ there is a difference. However, it only affects the $\Box R$
term, which is a trivial anomaly (it can be canceled by a
counterterm). Thus also in this case any $\rho$ is good enough
for the computation of the trace anomaly. To be specific, 
a $D=4$ conformal scalar needs a potential 
$V=\frac{\xi}{2}R$, where $\xi= {(D-2)\over 4(D-1)}={1\over 6}$.
Therefore $V_q$, which includes the DR counterterm, becomes 
$V_q= - {1\over 24} R$. 
Inserting $V_q$ into (\ref{zrho1}) 
produces the following trace anomaly for a conformal scalar field
\be
a_2 =
\left( -\frac{1}{144}C_\rho -\frac{1}{12}D_\rho
    +\frac{1}{24}E_\rho +\frac{1}{864} \right) \Box R 
    -\frac{1}{720}R_{\mu\nu}^2 +\frac{1}{720}R_{\mu\nu\a\b}^2 
\label{A4}
\ee
which is the correct value as far as the universal terms are concerned.
In fact, only the total derivative term has a $\rho$--dependent coefficient;
the well-known value $\frac{1}{720}$ is recovered with DBC.

$\bullet$ In~\cite{Kleinert:2003zq} it was argued that one could reach 
covariant results for the partition function density 
also using the ``string inspired'' propagator 
in arbitrary coordinates, i.e. through  
a noncovariant expansion.
However the method proposed in section 7 of~\cite{Kleinert:2003zq}
does not seem to be correct, as far as we understand it.
In particular, if one were to include an 
external potential $V$ in the path integral, there would be no terms in
that method that could covariantize it. In any case, 
these facts may leave some doubts about the correct 
covariantization obtained using RNC in the presence of external potentials.
To make sure that the previously described gauge-fixing in RNC
achieves the correct covariantization even with 
external potentials, in the next subsection 
we present a four-loop computation in RNC which tests this issue.

\subsection{A look at order $\beta^3$}
\label{section:6D-SDW-coeff}

In the last subsection it was pointed out that RNC, with the correct
factorization of the zero modes, should produce automatically the correct 
covariant result for the partition function density $z^{(\rho)}(x_0,\b)$,
also in the presence of terms that are  not explicitly constructed from 
the metric tensor such as an external scalar potential $V$. 
In fact, gauge independence of $Z(\beta)$  guarantees that 
$z^{(\rho)}(x_0,\b)$ should always contain a universal 
($\rho$-independent) covariant part and a $\rho$-dependent total derivative,  
which vanishes upon space-time integration. 
The $\rho$-dependent total derivative is in general 
a noncovariant expression (as seen in eq. (\ref{gntd}))
if calculated in arbitrary coordinates.
It is instead covariant if calculated with RNC.

To make sure that the correct covariantization 
happens in RNC also in the presence of an external potential $V$,
we compute the terms linear in $V$ 
that arise at order $\beta^3$. They belong to 
the Seeley--DeWitt coefficient $a_3$, and are enough to test
the correct covariantization of the result. This test is in the same spirit
of \cite{Schalm:1998ix}, which employed RNC to a higher perturbative order 
to discover problems in the computation of the trace anomaly
with the SI propagator.

Thus we analyze at order $\beta^3$ the partition function density
\be
z^{(\rho)}(x_0,\b) = 
\Big\la e^{-S_q^{(int)}}\Big\ra
\label{anomaly-V}
\ee
where, as before, we have included in $S^{(int)}_q$ an arbitrary external 
scalar potential $V_q= V -{1\over 8} R$.
It is enough to consider the terms linear in $V_q$. As a consequence,
we only  need  the perturbative  expansion  
of  the ghost action~(\ref{SFP}) to order $\beta^2$. We thus 
only need to calculate the correlators
\bea
&&\b \nabla_\a V_q \idts \lx[ {1\over 12\b}\nabla_{\a_1}R_{\a_2 \mu\nu\a_3} 
\Big\la \xi^\a(\t)\,\xi^{\a_1}\xi^{\a_2}\xi^{\a_3}
(\dot \xi^\mu \dot \xi^\nu +a^\mu a^\nu +b^\mu c^\nu)(\s)\Big\ra\rx.
\nonumber\\[.5mm] 
&&\hskip2cm\lx. +{1\over 12}\nabla_{\a_1}R_{\a_2\a_3} \Big\la \xi^\a(\t)\, 
\xi^{\a_1}\xi^{\a_2}\xi^{\a_3}(\s)\Big\ra 
\rho(\s)\rx]
\nonumber\\[.5mm]
&&+{\b \over 2}  \nabla_\a \nabla_\b V_q \idts \lx[ {1\over 6\b} 
R_{\a_1 \mu\nu\a_2} \Big\la \xi^\a\xi^\b(\t)\, 
\xi^{\a_1}\xi^{\a_2}(\dot \xi^\mu \dot \xi^\nu 
+a^\mu a^\nu +b^\mu c^\nu)(\s)\Big\ra\rx.
\nonumber\\[.5mm] 
&&\hskip2cm\lx. +{1\over 3} R_{\a_1\a_2} \Big\la \xi^\a\xi^\b(\t)\, 
\xi^{\a_1}\xi^{\a_2}(\s)\Big\ra \rho(\s)\rx]
\nonumber\\[.5mm]
&&
-{\b\over 4!}\nabla_{\a_1}\cdots\nabla_{\a_4}V_q
\idt\Big\la \xi^{\a_1}\cdots \xi^{\a_4} \Big\ra
-\b^3 V_q a_2^{(0)} 
\eea
where  $a_2^{(0)}$ is the  Seeley-DeWitt coefficient computed
with vanishing potential $V_q$.
It can be read off directly from~(\ref{zrho1}) 
\be
a_2^{(0)} = {1\over 720}(R_{\mu\nu\a\b}^2-R_{\mu\nu}^2)
+{1\over 1152} R^2+\lx(
{1\over 72}C_\r-{1\over 12}D_\r +{1\over 24} E_\r
-{1\over 1728} \rx)\Box R~. 
\ee
One thus gets for the relevant terms we wish to consider
\bea
&&{\b^3\over 6}\nabla_\mu V_q\, \nabla^\mu R (J_1 -J_2+J_3) 
+{\b^3\over 6}R_{\mu\nu}\,\nabla^\mu\nabla^\nu V_q
(J_4+J_5-2J_6+2J_7)
\nonumber\\[.5mm]
&&+{\b^3\over 12} R\, \Box V_q\,  J_9 (J_{11}-J_{12}+2J_{10})
-\b^3 \lx[ {1\over 8}\Box^2 V_q 
+{1\over 12}\nabla_\mu\lx(R^{\mu\nu}\nabla_\nu V_q\rx)\rx] J_8
\nonumber\\[.5mm]
&&
+{\b^3\over 6}V_q\, \Box R\, \lx(-{1\over 12}C_\r+{1\over  2}D_\r
-{1\over 4}E_\r+{1\over  288}\rx) +\cdots
\eea
where $J_i$ are the integrals reported and calculated
in appendix~\ref{sub:integrals}. We finally obtain
\bea
z^{(\rho)}(x_0,\b)
=1+\cdots+\beta^3 \Big (
\nabla_\mu\,Q^\mu -{1\over 480}\nabla_\mu\, V_q \nabla^\mu R+\cdots\Big )
+\cdots
\label{A-V}
\eea  
where  
\bea 
Q^\mu  &\equiv&  -{1\over 48}\lx(C_\r-{1\over  12}\rx)
  R\,\nabla^\mu  V_q  +{1\over 6}\lx(-{1\over  2}C_\r^2+{1\over  4}C_\r
  +2D_\r -{1\over 160}\rx)  R^{\mu\nu}\,\nabla_\nu    V_q    
  \nonumber\\[.5mm]    
&& +{1\over 6}\lx(-{1\over 12}C_\r+{1\over  2}D_\r-{1\over 4}E_\r
  +{1\over  288}\rx)V_q\,\nabla^\mu R
\nonumber\\[.5mm]   
&& -{1\over 8}\lx(C_\r^2-{5\over 18}C_\r-{2\over 3}E_\r+{1\over144}\rx)
  \nabla^\mu\,\Box V_q 
\eea 
is the ``current'' whose divergence gives the total  derivative term.

It is then clear  that~(\ref{A-V}) yields an unambiguous  
$\rho$-independent partition function $Z(\beta)$.
For $\rho(\tau)=\delta(\tau)$ one has the DBC propagator which yields  
directly the correct terms belonging to the local Seeley--DeWitt  
coefficient $a_3$ contained in $z^{(\rho)}(x_0,\b)$ \cite{Gilkey:iq}. 
Others $\rho$'s must then yield the same local result up to a $\rho$-dependent 
covariant total derivative\footnote{The coefficient $a_3$ is related 
to the trace anomalies in 6 dimensions \cite{Bastianelli:2000hi}, 
which have been calculated also using the quantum mechanical path integral 
with DBC in \cite{Bastianelli:2000dw}.}.
Thus, we tested successfully the use of RNC to obtain the {\em correct}
covariant results for $Z(\beta)$ also in the presence of external potentials.

\section{Supersymmetric nonlinear sigma models}

The $N=1$ supersymmetric nonlinear sigma model is relevant for 
the worldline description of a spin 1/2 particle coupled to gravity. 
We have described this worldline approach in \cite{Bastianelli:2002qw},
where the representation of the one-loop effective action was written
in terms of a path integral with periodic boundary conditions (PBC) for 
the bosonic coordinates $x^\mu$ and antiperiodic boundary conditions 
(ABC) for their supersymmetric partners, the worldline fermions 
$\psi^\mu$.
The worldline fermions play the role of the gamma matrices of the Dirac 
equation and the ABC are necessary to compute the trace in the spinorial 
space on which the gamma matrices act.
These boundary conditions break worldline supersymmetry.
In particular, the antiperiodic fermions $\psi^\mu$ do not have any zero 
mode. The zero modes of the bosons can then be treated as in the previous 
section. 

On the other hand, PBC for both bosons and fermions preserves 
supersymmetry. The change of boundary conditions from ABC to PBC
for the fermions corresponds to an insertion of the 
chiral matrix $\gamma^5$ inside the trace in spinorial space. This gives 
directly the regulated expression of the Witten index for the nonlinear 
sigma model \cite{Witten:df}, which is identified with the chiral $U(1)$
anomaly corresponding to a massless Dirac fermion coupled to gravity 
\cite{Alvarez-Gaume:1983at}.
In the present case, the periodic fermions acquire zero modes as well,
and it is of interest to consider their factorization.
In this section we describe this factorization using directly the action 
written in components. Superspace methods may also be used, but they do not 
seem to bring in drastic simplifications.

The $N=1$ supersymmetric nonlinear sigma model is described by the action
\bea 
S[x,\psi] = {1\over \beta}
\int_0^1 \!\! d\tau\, 
{1\over 2}g_{\mu\nu}(x) \biggl (\dot x^\mu \dot x^\nu 
+ \psi^\mu {D \psi^\nu \over d\tau} \biggr ) 
\label{sac}
\eea 
where 
${D \psi^\nu \over d\tau} = \dot \psi^\nu +\dot x^\lambda 
\Gamma_{\lambda\rho}^\nu(x) \psi^\rho$
is the worldline time derivative covariantized with the target space 
connection.  We consider PBC for both bosons and fermions,
$x^\mu(0)=  x^\mu(1)$,  $\psi^\mu(0) =\psi^\mu(1)$.
The corresponding partition function (Witten index) is then
\be
I_W= \oint {\cal D}x {\cal D}\psi \  e^{-S[x,\psi]}  \ .
\ee

In the flat limit ($g_{\mu\nu}=\delta_{\mu\nu}$) zero modes appear for 
both $x^\mu$ and $\psi^\mu$, corresponding to the invariances 
$\delta x^\mu(\tau) = \epsilon^\mu$ and $\delta \psi^\mu(\tau)=\theta^\mu$ 
for constant $\epsilon^\mu$ and $\theta^\mu$. 
In curved target space these two invariances are generically broken.
Nevertheless we wish to factorize these ``would be'' zero modes to be 
able to carry out a perturbative evaluation and at the same time present 
the partition function as an integral over them.
Thus we proceed as in section 2, and introduce extra dynamical 
gauge variables together with suitable gauge fixing conditions.
The introduction of extra gauge variables can be obtained by 
considering the following identities 
\bea
S[x,\psi] &=& S[x_0+y,\psi] = S[x_0+y(x_0,\xi),\psi(\tilde \psi)] 
\nonumber\\[.5mm] 
&=& S[x_0+y (x_0,\xi),\psi(\psi_0 + \chi)] \equiv S[x_0,\psi_0,\xi, \chi] 
\eea
where we first introduce the new gauge variable $x^\mu_0$, then change 
coordinates to RNC centered at $x^\mu_0$ (this change to new coordinates 
$\xi^\mu$ and $\tilde \psi^\mu$  is specified by the functions
$y^\mu(x_0^\nu,\xi^\nu)$ and $\psi^\mu(\tilde\psi^\nu)$), and finally 
introduce a new gauge variable $\psi^\mu_0$ by setting 
$\tilde\psi^\mu = \psi_0^\mu +\chi^\mu$.
We end up with the action $S[x_0,\psi_0,\xi, \chi]$ for the sigma model
in RNC 
\bea 
S[x_0,\psi_0,\xi, \chi]= {1\over \beta}
\int_0^1 \!\! d\tau\, 
{1\over 2 }g_{\mu\nu} (x_0, \xi) 
\biggl (\dot \xi^\mu \dot \xi^\nu 
+ (\psi_0^\mu + \chi^\mu)
{D (\psi_0^\nu + \chi^\nu)\over d\tau} 
\biggr ) 
\label{sac11}
\eea 
where the metric $g_{\mu\nu} (x_0, \xi)$ in RNC is given in (\ref{two}).
This action contains the desired shift gauge symmetries
encoded in the following BRST symmetry, 
suitably extended to two pairs of nonminimal auxiliary fields
\be
\begin{array}{ll}
\delta x^\mu_0 = \eta^\mu \Lambda 
&\quad\quad
\delta \xi^\mu(\tau)= -Q^\mu{}_\nu(x_0,\xi(\tau))\, \eta^\nu \Lambda \\[2mm]
\delta \psi^\mu_0 = \gamma^\mu \Lambda 
&\quad\quad
\delta \chi^\mu (\tau) = - \gamma^\nu \Lambda \\[2mm]
\delta \eta^\mu = 0  
&\quad\quad
\delta \gamma^\mu = 0 \\ [2mm]
\delta \bar \eta_\mu = i \pi_\mu \Lambda 
&\quad\quad
\delta \pi_\mu = 0  \\[2mm]
\delta \bar \gamma_\mu = i p_\mu \Lambda 
&\quad\quad
\delta p_\mu =0  \ .
\end{array}
\ee
The nonlinear bosonic gauge symmetry acting on $(x_0^\mu, \xi^\mu)$ 
is just as in the previous section, while the fermionic symmetry  
acts linearly on the $(\psi^\mu_0,\chi^\mu)$ fields.
We gauge fix by choosing the gauge fermion 
\be
\Psi = \bar \eta_\mu \int_0^1 \!\! d\tau \,\rho(\tau ) \xi^\mu(\tau ) +
\bar \gamma_\mu \int_0^1 \!\! d\tau \,\rho(\tau ) \chi^\mu(\tau )
\ee
parametrized by the single function $\rho(\tau)$ normalized to 
$\int_0^1 d\tau \rho(\tau)=1$.
This gauge fixing is covariant under reparametrization of $x_0^\mu$,
and yields the following gauge fixed action 
\bea 
&&S_{gf} [x_0,\psi_0,\xi, \chi,\eta,\bar\eta,\gamma,\bar\gamma,\pi,p]
\equiv S[x_0,\psi_0,\xi, \chi] 
+ {\delta\over \delta \Lambda}\Psi \nonumber\\
&& = S[x_0,\psi_0,\xi, \chi] +  
i \pi_\mu \int_0^1 \!\! d\tau \,\rho(\tau ) \xi^\mu (\tau )-\bar\eta_\mu  
\int_0^1 \!\! d\tau \,\rho(\tau ) 
Q^\mu{}_\nu(x_0, \xi(\tau ))\, \eta^\nu\nonumber\\
&&\quad -i p_\mu \int_0^1 \!\! 
d\tau \,\rho(\tau ) \chi^\mu (\tau )-\bar\gamma_\mu  
\gamma^\mu  \ .
\label{ggf2}
\eea
The ghosts $\gamma^\mu,\bar \gamma_\mu$ can be trivially integrated out,
while integration over the lagrange multipliers $\pi_\mu$ and $p_\mu$
produces the constraints
\be
\idt \rho(\tau) \xi^\mu(\tau)=0 \ , \quad \quad 
\idt \rho(\tau) \chi^\mu(\tau)=0  \ .
\label{gcon2}
\ee
These constraints permit to  
invert the kinetic term to find the perturbative propagators
for the fields $\xi^\mu$ and $\chi^\mu$.
Thus we end up with the path integral 
\bea
I_W &=& \int\! dx_0\, d\psi_0\, d\eta\, d\bar\eta 
\oint\! D\xi\, D\chi\  
\delta \Big(\int\!\!\rho\,\xi^\mu\Big)\, 
\delta\Big(\int\!\!\rho\,\chi^\mu\Big)
\nonumber\\[.5mm]
&&\hskip2cm \times \exp \big(-S[x_0,\psi_0,\xi, \chi]
- S_{FP}[x_0,\xi,\eta,\bar\eta] \big)
\label{inwi}
\eea
where $S_{FP}$ is the same Faddeev-Popov action of
the bosonic model written in (\ref{SFP}).
Note that since we use fermionic fields 
with curved indices ($\psi_0^\mu$, $\chi^\mu$) their nontrivial
path integral measure is exactly compensated by the nontrivial measure 
of their bosonic supersymmetric partners ($x_0^\mu$, $\xi^\mu$). 
Therefore the complete measure 
is flat and no extra ghosts are needed in this case, as shown in   
\cite{Bastianelli:2002qw}.

The path integral can now be computed. Its value should not depend on 
$\beta$ as consequence of supersymmetry \cite{Witten:df}. 
Thus a semiclassical calculation produces already the complete
result \cite{Alvarez-Gaume:1983at}.
However, we have now set up the path integral in such a way that higher 
loop calculations can be unambiguously performed to test the 
$\beta$ independence. We perform this test in the next subsection.

\subsection{Order $\beta$ correction to the chiral anomaly}
\label{section:anomaly}

In this section we consider the perturbative expansion for the
$N=1$ supersymmetric sigma model with periodic boundary conditions 
on all fields, as described in the previous section.
We will explicitly verify the $\beta$ and $\rho$ independence of the path
integral for the chiral anomaly $I_W$ in $D=2$ and $D=4$ 
up to order $\beta$ (i.e. two loops on the worldline), employing dimensional 
regularization~\cite{Bastianelli:2002qw} whenever necessary. 
Before starting, let us recall that the chiral anomaly $I_W$ can also be 
written as
\bea
I_W = 
{\rm Tr}\left( (-1)^F \, e^{-\b Q^2} \right) =
{\rm Tr}\left( \g_5 \, e^{{\b\over 2}{\nabla\!\!\!\!/}^{\ 2}} \right) 
\label{Ad}
\eea
where $Q= {i\over \sqrt{2}}{\nabla\!\!\!\!/}$ is the supersymmetry charge 
(the Dirac operator), 
$H=Q^2=-{1\over 2}{\nabla\!\!\!\!/}^{\ 2}= - {1\over 2}\Box + {1\over 8}R$  
the quantum hamiltonian, and $(-1)^F=\gamma_5$ the fermion number operator
of the $N=1$ supersymmetric sigma model.
Dimensional regularization preserves supersymmetry without the need of any 
counterterm \cite{Bastianelli:2002qw}:
the quantum hamiltonian $H$ must have a potential ${1\over 8} R$,
obtained from squaring the supercharge $Q$, and DR produces exactly that
potential.
Thus the path integral representation of $I_W$ in (\ref{inwi}) is 
good as it stands when using dimensional regularization. Other 
regularization schemes may need counterterms to enforce supersymmetry.

Let us consider the perturbative expansion at fixed $x_0$ and  $\psi_0$.
One obtains the various propagators from the free action
\bea
S_2 = {1\over 2\beta}\, g_{\mu\nu}(x_0) \int_0^1 \!\! d\tau 
\left(\dot \xi^\mu \dot \xi^\nu + \chi^\mu \dot\chi^\nu\right)
-\bar\eta_\mu\eta^\mu
\label{action-free}
\eea
with the fields constrained by (\ref{gcon2}).
The propagators for $\xi^\mu$ and the ghosts $\eta^\mu, \bar\eta_\mu$ were
already described in (\ref{generic_prop}), whereas the one
for the fermions $\chi^\mu$ is given by
\be
\la \chi^\mu(\t)\chi^\nu(\s)\ra = \beta\, g^{\mu\nu}(x_0)\, \F(\t,\s)
\ee
where the Green function
\be
\F(\t,\s) = \ddel(\t-\s)-\ids\!\!' \, \rho(\s')\,\ddel(\tau-\s') +
\ids\!\!' \, \rho(\s')\,\ddel(\sigma-\s')
\label{generic_prop-fermi}
\ee
satisfies
\bea
\dF(\t,\s) &=& \delta(\t-\s)-\r(\t)   \nonumber\\[.5mm]
-\Fd(\t,\s) &=& \delta(\t-\s)-\r(\s) 
\eea
(dots on the left/right denote derivatives with respect to
the first/second variable) and
\bea
\idt \rho(\tau)\, \F(\t,\s)=0 \ .
\eea
With these propagators the chiral anomaly is given by
\bea
I_W= \int {d^D x_0\, d^D \psi_0\over (2\pi i)^{D\over 2}}
   \,\, \la e^{- S_q^{(int)}} \ra
\label{pi-PBC}
\eea
where $ S_q^{(int)}=S_q-S_{q,2}$, $S_q =  S+S_{FP}$ is the total action 
in (\ref{inwi}), and $\la 1\ra=1$ is the normalization of the 
($\r$-dependent) path integral. 
For the sake of compactness, it is useful to rescale 
$\psi_0 \to \sqrt{\b}\,\psi_0$, getting 
$d^D\psi_0 \to \b^{-D/2}\, d^D\psi_0$ and $\tilde\psi \to 
\sqrt{\b}\,\psi_0+\chi$.
Being $\tilde\psi$ of order $\sqrt{\b}$ we can organize the interaction 
terms in a systematic way, namely we split 
$S_q^{(int)}= S_{q,3}+ S_{q,4}+\ldots$ with the $S_{q,n}= S_n +S_{FP,n}$ 
contributions of order $O(\beta^{{n\over 2}-1})$.
Moreover, in the RNC expansion 
$ S_3=S_{FP,3}=0$, as $g_{\mu\nu,\l}(x_0)=0$. Thus, after some simplifying 
manipulations that exploit index symmetries, the relevant vertices for 
our purposes read\footnote{$S_{5,1}$ and  $S_{6,1}$ are 
contributions to $S_5$ and $S_6$ that come from the cubic 
and quartic order in the expansion of $g_{\mu\nu}(x)$ in 
RNC, see~(\ref{two}). They are not explicitly reported here 
since they will not enter in the forthcoming calculations.}
\bea
 S_4 &=& {1\over 6\beta} R_{\alpha\mu\nu\beta}\idt \xi^\a 
\xi^\b \biggl (\dot \xi^\mu \dot \xi^\nu
+(\sqrt{\beta}\,\psi_0^\mu+\chi^\mu)\dot \chi^\nu\biggr )
\nonumber\\[.5mm]
 &&+{1\over 4\beta}R_{\alpha\mu\nu\lambda}\idt 
\xi^\alpha \dot \xi^\mu (\sqrt{\b}\,\psi_0^\nu+\chi^\nu) 
(\sqrt{\b}\,\psi_0^\lambda+\chi^\lambda)
\label{s4}\\[2mm]
 S_5 &=& S_{5,1}
+{1\over 6\beta} \nabla_\beta R_{\alpha\mu\nu\lambda} \idt
\xi^\alpha \xi^\beta \dot \xi^\mu (\sqrt{\b}\,\psi_0^\nu+\chi^\nu) 
(\sqrt{\b}\,\psi_0^\lambda +\chi^\lambda)
\label{s5}\\[2mm]
 S_6 &=&   S_{6,1} +{1\over \beta}\left( {5\over 72} R_{\l\a\b\s}
R_{\gamma\mu\nu}{}^\s +{1\over 24}R_{\l\a\nu\s}R_{\b\mu\gamma}{}^\s
+{1\over 36}R_{\lambda \a\b\s}R_{\g\nu\mu}{}^\s \right.  
\nonumber\\[.5mm]
&& \left.+{1\over 16} \nabla_\gamma \nabla_\b R_{\a\mu\nu\l}\right) 
 \idt \xi^\a \xi^\b \xi^\gamma \dot \xi^\mu 
(\sqrt{\b}\,\psi_0^\nu+\chi^\nu) 
(\sqrt{\b}\,\psi_0^\lambda +\chi^\lambda)
\label{s6}
\eea
where all tensors are evaluated at the point $x_0$.

It is clear from the Grassmannian nature of the measure that the only 
terms that can give an {\em a priori} nonvanishing contribution to 
(\ref{pi-PBC}) are those that saturate the measure itself. Hence, in $D$ 
dimensions only terms with $D$ fermionic zero modes can contribute; in 
particular
\bea
\int d^D\psi_0\ \psi_0^{\mu_1}\cdots\psi_0^{\mu_D} = 
\varepsilon^{\mu_1 \cdots \mu_D}\ .
\eea
Before starting our calculation, let us briefly recall how some known 
results for chiral anomalies in arbitrary dimensions $D$ 
can be derived from (\ref{pi-PBC}). First of all, the 
chiral anomaly is trivially zero when $D=2k-1,\, k\in {\mathbb N}$,
because an odd
number of $\psi_0$ is needed to saturate the measure whereas the fermions
always come in pairs. Conversely, if one considers $D=2k$, it is not 
difficult to convince oneself~\cite{Alvarez-Gaume:1983at, 
Hatzinikitas:1997md} that, at the one-loop level, the only
nonvanishing contributions come from the $k$-th power of the vertex 
${1\over 4}R_{\a\mu\nu\l} \int_0^1\!d\t\, \xi^\a \dot \xi^\mu 
\psi_0^\nu \psi_0^\l$~; hence, no fermionic propagators are involved. More 
specifically, the independent terms contributing to the anomalies are of 
the form
\bea
{\rm Tr} \left[{\cal R}^k \right] L_k\ , \  \ \
{\cal R}_{\a\b}\equiv R_{\alpha\beta\mu\nu}\, \psi_0^\mu\psi_0^\nu
\eea
where the integrals $L_k$ are given by
\bea
L_k = \int_0^1 \!\!\! d\t_1\cdots\int_0^1 \!\!\! d\t_k \; 
\dB(\t_1,\t_2)\dB(\t_2,\t_3)\cdots
\dB(\t_k,\t_1)\ .
\eea
Integrating by parts and making use of the identity 
$\dBd = \ddeld$, the latter can be reduced to their string-inspired 
counterparts
\bea
L_k &=& \int_0^1 \!\!\! d\t_1\cdots\int_0^1 \!\!\! d\t_k \; 
\ddel(\t_1 -\t_2)\ddel(\t_2-\t_3)\cdots
\ddel(\t_k-\t_1) \nonumber\\[2mm]
&=&
\left\{
\begin{array}{ll} 
{2\over (2\pi i)^{2l}}\, \zeta(2l) &\ \ k=2l \\[2mm]
0 &\ \ k=2l+1
\end{array}\right.  \qquad l \in \mathbb{N} 
\label{Jk}
\eea  
where $\zeta$ is the Riemann zeta function.
Notice that for $k$ odd, both traces and integrals vanish: 
indeed, gravitational contributions to the chiral $U(1)$ anomaly 
are present only for spacetime dimensions $D=4n, \, n\in {\mathbb N}$. 
The result of this semiclassical 
approximation is known to be exact~\cite{Alvarez-Gaume:1983at}, and it is 
independent of the boundary conditions chosen for the quantum
fluctuations (see also~\cite{Hatzinikitas:1997md}). 
Nevertheless, the full path integral (\ref{inwi}) should be independent 
of $\beta$ and $\rho$. We have explicitly tested this property by 
doing a two-loop calculation for the simplest cases $D=2,4$. 
For $D=2$ each worldline graph vanishes, so the proof is 
quite trivial.  Thus, we directly pass to describe the $D=4$ case.

The two-loop truncation of~(\ref{pi-PBC}) reads
\bea
I_W = -\int {d^4x_0\, d^4\psi_0\over (2\pi\beta)^2}\, \exp\left[ 
\left\la - S_{q,4}- S_{q,6}+{1\over 2} ( S_{q,4}^2+ S_{q,5}^2) + 
S_{q,4} S_{q,6} -{1\over 3!} S_{q,4}^3 \right\ra_c \, \right] \nonumber\\
\label{anom}
\eea  
where we have omitted those terms that are explicitly zero; 
the subscript $c$ refers to connected diagrams. To further simplify 
this expression, let us make few preliminary considerations. 
Recalling the splitting $S_q=S+S_{FP}$, we compute
\bea
\label{s4-vev}
\la  S_4\ra &=& {1\over 6\beta} R_{\alpha\mu\nu\beta}\idt 
\la \xi^\alpha \xi^\beta(\dot \xi^\mu \dot \xi^\nu
+(\sqrt{\b}\,\psi_0^\mu+\chi^\mu)\dot \chi^\nu)\ra\nonumber\\[.5mm] 
&&+{1\over 4\beta}R_{\alpha\mu\nu\lambda}\idt 
\la \xi^\alpha \dot \xi^\mu (\sqrt{\b}\,\psi_0^\nu+\chi^\nu)
(\sqrt{\b}\,\psi_0^\lambda+\chi^\lambda)\ra
\nonumber\\[2mm] &=& 
{\beta\over 6} R_{\alpha\mu\nu\beta}\idt \left[ g^{\alpha\beta} g^{\mu\nu} 
\B|_\t (\dBd+\dF)|_\t
+ g^{\alpha\nu} g^{\beta\mu}(\dB|_\t)^2\right]
\nonumber\\[2mm]
&=& -{\beta\over 2}\, C_\r\, R~, 
\eea
where we have used some of the formulae in appendix~\ref{sub:conventions}. 
This term does not contain fermionic zero modes.
This is also true for all the vertices coming from the ghost action $S_{FP}$. 
We also note  that $\la  S_6\ra$ is of order
$\beta^2$ and contains at most two zero modes. 
This implies that $S_6$ can only enter at two-loops
through connected terms. 
Thus, the (one-loop) anomaly is given by 
\bea
I_W^{(0)} &=& -\int {d^4x_0\, d^4\psi_0\over (2\pi\beta)^2}\, 
{1\over 2}\left\la   S_4^2
\right\ra_c
=\int {d^4x_0\, d^4\psi_0\over (2\pi)^2}\,{2\over 2\cdot 4^2}\, 
\psi_0^{\alpha_1}\cdots\psi_0^{\alpha_4}\, 
R_{\alpha_1\alpha_2\mu\nu}\, R_{\alpha_3\alpha_4}{}^{\mu\nu}\, L_2
\nonumber\\[2mm]
&=& -{1\over 768\pi^2}\int d^4x_0\ \varepsilon^{\alpha_1\cdots\alpha_4}
R_{\alpha_1\alpha_2\mu\nu}\, R_{\alpha_3\alpha_4}{}^{\mu\nu}~, 
\label{anomaly:one-loop}
\eea
where from eq. (\ref{Jk}) $L_2 = - {1\over 12}$, as expected. 

From these preliminaries one recognizes that
the order $\beta$ correction is given by
\bea
I_W^{(1)} &=& \int {d^4x_0\, d^4\psi_0\over (2\pi\beta)^2}\, 
\left\{ {1\over 2}\la   S_4\ra  \la   S_4^2\ra_c 
-{1\over 2}\la   S_5^2\ra_c
-\la  S_4   S_6\ra_c +{1\over 3!}\la   S_4^3\ra_c 
\right.\nonumber\\&&\left.\hskip3cm \hspace{-2ex}
+{1\over 2}\la S_{FP,4}\ra \la  S_4^2\ra_c
+{1\over 2}\la S_{FP,4}\,  S_4^2\ra_c \right\}~.
\label{anomaly:two-loop}
\eea 
We first compute the  four terms in the first line, leaving 
the contributions from the FP vertices in the second line for the very end.

The first term can be immediately
obtained from~(\ref{s4-vev}) and (\ref{anomaly:one-loop}). 
Using the identities listed in appendix~\ref{sub:conventions}, it
can be written as
\bea
{1\over 2}\int {d^4x_0\, d^4\psi_0\over (2\pi\beta)^2}\,\la   S_4\ra \la 
  S_4^2\ra_c =
-{\b\over 384}\int {d ^4x_0\over (2\pi)^2}\, C_\r\, (2 K_0-8K_2)~.
\eea
As for the second term, the only way to pick out four zero 
modes is to square the connection part, the one explicitly reported
in~(\ref{s5}), obtaining
\bea
\left\la  S_5^2\right\ra_c^{(4\psi_0)} &=& {1\over 36}\, 
\nabla_{\beta_1} R_{\alpha_1\mu_1\lambda_1\lambda_2}\,
\nabla_{\beta_2} R_{\alpha_2\mu_2\lambda_3\lambda_4}\, 
\psi_0^{\lambda_1}\cdots\psi_0^{\lambda_4}\nonumber\\[.5mm]
&&\hskip1cm\times \int_0^1\!\!d\t_1\int_0^1\!\!d\t_2
\left\la \xi^{\a_1}\xi^{\b_1}\dot \xi^{\mu_1}\ \xi^{\a_2}\xi^{\b_2}
\dot\xi^{\mu_2}\right\ra\nonumber\\[2mm]
&=& {\beta^3\over 36}\, \psi_0^{\lambda_1}\cdots\psi_0^{\lambda_4}
\left[ \nabla_{\alpha} R^\alpha{}_{\s\lambda_1\lambda_2}\nabla_{\beta} 
R^{\beta\s}{}_{\lambda_3\lambda_4}(2I_2 - I_3 - I_4)
\right.\nonumber\\[.5mm]
&&\left.\hskip1cm +\nabla_{\a} R_{\b\s\lambda_1\lambda_2}(\nabla^{\b} 
R^{\alpha\s}{}_{\lambda_3\lambda_4}+
\nabla^{\alpha} R^{\beta\s}{}_{\lambda_3\lambda_4})(I_5-I_6)\right] \ .
\eea
Using the identities in appendix~\ref{sub:conventions} 
it becomes
\bea
-{1\over 2}\int {d^4x_0\, d^4\psi_0\over (2\pi\beta)^2}\, \left\la 
  S_5^2\right\ra_c &=&
-{\beta\over 2}\int {d ^4x_0\over (2\pi)^2}\left[-{1\over 72} 
\lx( K_0+4K_1\rx)\,(2I_2 - I_3 - I_4)  
\right.\nonumber\\[.5mm] 
&& \left. \hskip-1.5cm +{1\over 18}K_2\, \left(4I_2-2I_3-2I_4+3I_5
 -3I_6\right)+{1\over 6} K_3\, \left(I_6-I_5\right)\right]~. \quad\quad
\label{s52:final}
\eea
The third term yields
\bea
\left \la   S_4   S_6\right\ra_c^{(4\psi_0)} &=& {1\over 4}\,
 R_{\beta_1\mu_1\alpha_1\alpha_2}\, 
R_{\r_2\s_2\gamma_2\xi_2\alpha_3\alpha_4}
\, \psi_0^{\alpha_1}\cdots\psi_0^{\alpha_4}\nonumber\\[.5mm]
&&\hskip1cm\times
\int_0^1\!\!d\t_1 \int_0^1\!\!d\t_2\, \la \xi^{\beta_1}\dot \xi^{\mu_1}\ 
\xi^{\r_2}\xi^{\s_2}\xi^{\gamma_2}\dot \xi^{\xi_2}\ra
\nonumber\\[2mm]
&=& -{\beta^3\over 2}\,R_{\beta_1\mu_1\alpha_1\alpha_2}\, 
R_{\r_2\s_2\gamma_2\xi_2\alpha_3\alpha_4}
\, \psi_0^{\alpha_1}\cdots\psi_0^{\alpha_4}\nonumber\\[.5mm]
&&\hskip1cm\times g^{\mu_1\xi_2}(g^{\r_2\s_2}g^{\beta_1\gamma_2}+ 
g^{\r_2\gamma_2}g^{\beta_1\s_2}
+g^{\s_2\gamma_2}g^{\beta_1\r_2})\, I_1
\eea
where the six-index tensor is the one given in the round parenthesis of 
equation~(\ref{s6}). After a bit of tensor algebra one gets
\bea
- \int {d^4x_0\, d^4\psi_0\over (2\pi\beta)^2}\, \left\la   S_4 
  S_6\right\ra_c 
&=& {\beta\over 2}\int{d^4 x_0\over (2\pi)^2}
  \left(-{1\over 24}K_0 +{2\over 9}K_1-{5\over 9}K_2
  +{1\over3}K_3\right) I_1 \ . \quad  
\eea
Finally, we are left with the fourth and more involved term
\bea
\left \la   S_4^3\right\ra_c^{(4\psi_0)} &=& \left\la \left( S_{4,1}
+ S_{4,2}\right)^3\right\ra_c^{(4\psi_0)}
\label{s43}
\eea
where $ S_{4,1}$ and $ S_{4,2}$ can be read off from~(\ref{s4})
as the first and second term, respectively.
The former may yield only one
zero mode and thus it does not give any contribution 
to~(\ref{s43}) by itself. The remaining terms give instead
\bea
{3\over 3!}\int {d^4x_0\, d^4\psi_0\over (2\pi\beta)^2}\,
\left\la ( S_{4,1})^2  S_{4,2}\right\ra_c &=& 
-{\beta\over 144}\int{d^4 x_0\over (2\pi)^2} 
\,(4K_2-K_3)\, I_7\\[2mm]
{3\over 3!}\int {d^4x_0\, d^4\psi_0\over (2\pi\beta)^2}\,
\left\la   S_{4,1} ( S_{4,2})^2\right\ra_c &=& 
- \beta\int{d^4 x_0\over (2\pi)^2}
\left[ {1\over 48}K_0\, (3 I_{14}+I_8)\right.\\[.5mm]
&&+{1\over 24}K_1\, (-2I_{11}+4I_{13}-2I_{12}-I_{10}-I_9)
\nonumber\\[.5mm]
&&\left.+{1\over 72}K_2\, (6I_{11}-12I_{13}+6I_{12}
-6I_8+3I_{10}+3I_9)\right]\nonumber\\[2mm]
{1\over 3!}\int {d^4x_0\, d^4\psi_0\over (2\pi\beta)^2}\,
\left\la ( S_{4,2})^3\right\ra_c &=& 
{\beta\over 16}\int{d^4 x_0\over (2\pi)^2}
\, K_3\, (2I_{15}-I_{16}-I_{17})~.
\eea  

We now explicitly use the integrals listed in appendix \ref{sub:integrals}, 
so that the four terms computed so far sum up to 
\bea
&& -\beta\int{d^4 x_0\over (2\pi)^2}\left[ 
{1\over 288}\lx(C_\r+{1\over 12}\rx)\, K_0
+{1\over 72}\left(-{7\over 36}C_\r -{5\over 3}E_\r 
+{1\over 720}\right)K_1\right.\nonumber\\[.5mm]
&&\left.\hspace{2ex} +{1\over 72}\left(-{19\over 36}C_\r 
+{7\over 3}E_\r -{7\over 144}\right)K_2
+{1\over 72}\left({5\over 36}C_\r +{1\over 3}E_\r 
+{13\over 720}\right)K_3\right]~. \quad
\eea 
This is non zero unless one adopts DBC Feynman rules, for which 
$C_{\rm DBC} = -{1\over 12}$, $E_{\rm DBC} = {1\over 180}$, so that 
$I_W^{(1)} \propto {\displaystyle \int\sum_{l=1}^3} K_l =0$
by Bianchi identities.  
However, the Faddeev-Popov action gives the additional contributions
\bea
&&{1\over 2}\int {d^4x_0\, d^4\psi_0\over (2\pi\beta)^2}\,
\left[ \left\la S_{FP,4}\right\ra \left\la (  S_{4,2})^2\right\ra_c 
+\left\la S_{FP,4}\,  (  S_{4,2})^2\right\ra_c\right]
\nonumber\\[.5mm]
&&= \beta \int{d^4 x_0\over (2\pi)^2}\left[ {1\over 288} \lx(C_\r+
{1\over 12}\rx)\, 
(K_0-4K_2)+{1\over 6}(K_2-K_1)\, I_{18}\right]
\eea
which lead to the complete order-$\beta$ correction 
\bea
I_W^{(1)} = -\beta\int{d^4 x_0\over (2\pi)^2}{1\over 72}
\left({5\over 36}C_\r +{1\over 3}E_\r +{13\over 720}\right)
\sum_{l=1}^3 K_l &=& 0 \ .
\eea
This result shows that one may safely 
adopt $\r$-dependent Feynman rules also in the case of periodic worldline 
fermions, provided that one appropriately takes care of the zero modes. 
As we showed, this can be done through the 
appropriate introduction of auxiliary gauge variables,
together with a suitable gauge fixing of the ensuing shift gauge symmetry,
which turns the auxiliary gauge variables into the wanted zero modes.
As a consequence of this procedure, one must include 
the corresponding FP determinant, which vanishes only in the DBC case. 
Indeed, in \cite{Waldron:1995tq}, where a similar calculation has been 
carried out in DBC as a test for the time slicing regularization, no
extra vertices were needed to obtain the correct result.

\section{Conclusions}

We have discussed in great details the factorization of 
perturbative zero modes which appear in the worldline approach to one-loop 
quantities for spin 0 and spin 1/2 particles.
We have focused particularly on the issue of reparametrization invariance 
of the corresponding nonlinear sigma models.
As well-known, the calculation of one-loop quantities in the worldline 
formalism requires to master the path integral for sigma models
on the circle.
In perturbative computations around flat target space, 
perturbative zero modes appear. We have described their factorization
using BRST methods and employed an arbitrary function $\rho(\tau)$,
the so-called background charge, to parametrize different gauges.
The arbitrary function $\rho(\tau)$ allows to test for gauge independence
and includes as particular cases the two most used methods:
$i)$ The method based on Dirichlet boundary conditions (DBC),
which is directly related to the calculation of the heat kernel. It 
produces covariant results in arbitrary coordinates
also before integrating over zero modes, but it has a propagator 
that is not translational invariant on the worldline.
$ii)$ The ``string inspired'' method, where the constant 
modes in the Fourier expansion of the quantum fields
are factorized. It yields a translational invariant propagator,
but when using arbitrary coordinates produces noncovariant total derivatives 
in the partition function density. The string inspired propagator
is simpler and faster for computations on the circle,
and the technical problem of the noncovariant total derivatives 
can be overcome by using Riemann normal coordinates.
 
The BRST symmetry guarantees the gauge independence of the partition 
function $Z(\beta)$. From this one deduces that the partition function 
density $z^{(\rho)}(x_0,\b)$ should contain a universal 
$\rho$-independent covariant part and at most a $\rho$-dependent total 
derivative, which vanishes upon space-time integration. 
The $\rho$-dependent total derivative is in general a noncovariant expression 
when calculated in arbitrary coordinates, as exemplified in eq. (\ref{gntd}).
However, it is covariant if calculated using Riemann normal coordinates.
The source of this discrepancy is easy to understand: the gauge-fixing
constraint is expressed in terms of coordinate differences $x^\mu - x_0^\mu$
and does not transform covariantly as a vector under reparametrization of the
background point $x_0^\mu$, but it does once one chooses 
Riemann coordinates centered at $x_0^\mu$.
The gauge fixing in Riemann coordinates is rather subtle.
It has originally been studied by Friedan for applications to bosonic 
2D nonlinear sigma models \cite{Friedan:1980jm}, and recently employed by 
Kleinert and Chervyakov \cite{Kleinert:2003zq} in the 1D case.
It makes use of a suitable nonlinear shift symmetry.

In this paper we have first addressed some issues left open 
in \cite{Kleinert:2003zq}, and performed extensive tests
to check the correctness of the BRST method for extracting 
the zero modes. Then we have extended
this method to the supersymmetric nonlinear sigma model, which is
relevant to the description of spin 1/2 particles coupled to gravity.
All our tests have been successful. We can now draw the following
conclusions:

$\bullet$ 
The perturbative calculation of $Z(\beta)$ is gauge independent if
the external fields describing the interactions have derivatives
vanishing sufficiently fast at infinity. 
This is a necessary condition for the validity of the perturbative
expansions employed in our tests. If these conditions are met,
all total derivatives integrate to zero.
When the external fields have derivatives which do not 
vanish fast enough, one may still proceed for example by trying to resum 
a class of diagrams and  make the perturbative expansion of $Z(\beta)$ 
well-defined.
This works in the harmonic oscillator, which has the potential
$V(x) = {1\over 2}\omega^2 x^2$. In this case we have explicitly checked 
the gauge independence of the partition function up to order 
$(\beta \omega)^2$.

$\bullet$ 
The covariant local expansions of $z^{(\rho)}(x_0,\b)$ in Riemann normal 
coordinates (RNC) are different for different $\rho$'s. 
The difference is given by covariant total derivatives with coefficients 
depending on $\rho$.

$\bullet$
One can use RNC with any $\rho$ to compute local trace anomalies in $D=4$.
In particular, one can use the string inspired propagator. 
The total derivatives only affect 
the $\Box R$ term, which anyway is a trivial anomaly 
(it can be canceled by a counterterm). 
The reason why the ``string inspired''
computation tested in \cite{Schalm:1998ix} failed is 
because the Faddeev-Popov contribution arising form the nonlinear shift 
symmetry needed for the implementation of RNC was missing. 

$\bullet$ The use of RNC yields local covariant results also in the presence
of external fields like scalar and vector potentials.
We have explicitly verified this in the presence of a scalar potential $V$.
We have performed a higher loop calculation to make sure that the
expected results are indeed obtained.
It is not obvious how to use arbitrary coordinates to check 
this property. 
In particular, the method proposed in section 7 of~\cite{Kleinert:2003zq}
to achieve a local covariant results in arbitrary coordinates 
does not seem to be correct, as for example it does not yield the correct 
covariantization of the higher derivatives of the scalar potential.

$\bullet$ 
We have used the supersymmetric model with the correct factorization
of the zero modes to test successfully for any $\rho$ 
the $\beta$ independence of the Witten index, i.e. of the chiral anomaly 
for a spin 1/2 particle.
We have computed the order $\beta$ corrections 
the chiral anomaly in $D=2$ and $D=4$ and found that they vanish.

The results described here make sure that
path integrals for nonlinear sigma model on the circle, 
with and without supersymmetry, are in good shape both 
in their UV and IR structure. 
They can be employed to produce unambiguous results in the worldline 
formalism with background gravity. 
The method discussed here could also turn out to be 
relevant for the extension to the noncommutative case,
recently studied 
in \cite{deBoer:2003cp} for describing the coupling of 
$D0$ branes to gravity, and to the free field analysis of the AdS/CFT
correspondence \cite{Gopakumar:2003ns}.

\acknowledgments{The authors thank the EC Commission for financial
support via the FP5 Grant HPRN-CT-2002-00325}. OC would also like to
thank the CPHT of the \'Ecole Polytechnique for their 
kind hospitality while parts of this work were completed.
The work of OC has been partially supported by a Marco Polo fellowship 
of the Universit\`a di Bologna.

\appendix
\section{Appendix}
\label{section:appendix}

\subsection{Linear sigma model}
\label{sub:linear-sigma}
Let us consider the linear sigma model
\be 
S={1\over \beta}
 \idt \left [
{1\over 2}\delta_{\mu\nu}\dot x^\mu \dot x^\nu +\beta^2 V(x) \right ] 
\ee
and compute perturbatively in $\beta$ the partition function
\be
Z = {\rm Tr}\ e^{-\beta H}= \oint Dx\  e^{-S[x]}  \ .
\label{A.pf}
\ee
We first try to understand the factorization of the zero mode
with some generality. 
To extract the zero modes one can first introduce a shift symmetry, 
and then gauge fix it in a suitable way. Using the Faddeev--Popov method
one obtains the following chain of identities
\bea 
 Z &=& \oint Dx \ e^{-S[x]} \nonumber\\[.5mm]
&=& 
 \oint Dy\  e^{-S[x_0 +y]} \nonumber\\[.5mm]
&=& 
 \oint Dy \int d^D\epsilon\  \delta 
\Big (\int d\tau\, \rho(\tau)  y^\mu_\epsilon(\tau)  \Big )\, 
e^{-S[x_0 +y]} \nonumber\\[.5mm]
&=& 
 \int d^D\epsilon\,  
\oint Dy_\epsilon \   
\delta \Big (\int d\tau\, \rho(\tau) y^\mu_\epsilon(\tau)  \Big)\,
e^{-S[x_{0\epsilon} +y_\epsilon ]} \nonumber\\[.5mm]
&=& 
\int d^Dx_{0\epsilon}\,   \oint Dy\  
\delta \Big (\int d\tau\, \rho(\tau) y^\mu(\tau)  \Big)\,
e^{-S[x_{0\epsilon} +y ]} \nonumber\\[.5mm]
&=&
\int d^Dx_0 \, \oint Dy\  
\delta \Big (\int d\tau\, \rho(\tau) y^\mu(\tau)  \Big)\,
e^{-S[x_{0} +y ]} 
\equiv
\int {d^Dx_0 \over (2\pi\beta)^{D\over 2}}
\, z^{(\rho)}(x_0) \ .
\label{gfac}
\eea
Here we have first used the linear split $x^\mu(\tau)=x^\mu_0+y^\mu(\tau)$ 
for an arbitrary constant $x^\mu_0$,
and then the  translation invariance of the path integral measure
$Dx=Dy$. Then we made use of the shift invariance 
(typical of the background field method)
\bea
x^\mu_0 &\to& x^\mu_{0\epsilon}=x^\mu_0+\epsilon^\mu\nonumber\\[.5mm]
y^\mu &\to& y^\mu_\epsilon=y^\mu-\epsilon^\mu
\label{shift}
\eea
which leaves invariant the field $x^\mu(\tau)$ since
$x^\mu(\tau)=x^\mu_0+y^\mu(\tau) =x^\mu_{0\epsilon}+y^\mu_\epsilon(\tau)$.
Therefore the action $S[x]$ itself remains invariant.
One may then use the Faddeev--Popov trick of inserting unity
represented as follows
\bea
1&=&\int d^D\epsilon \ 
\delta \Big (\int d\tau\, \rho(\tau) y^\mu_\epsilon(\tau)  \Big)
\nonumber\\[.5mm]
&=& 
\int d^D\epsilon  \
\delta \Big (\int d\tau\, \rho(\tau) (y^\mu(\tau) -\epsilon^\mu)\Big)
\nonumber\\[.5mm]
&=&
\int d^D\epsilon \ 
\delta \Big ( \int d\tau\, \rho(\tau) y^\mu(\tau) -\epsilon^\mu\Big)
\label{gf}
\eea
where the background charge is normalized 
to unity, $ \int_0^1 d\tau\, \rho(\tau) =1$.
Finally we used that $ d^D\epsilon  = d^Dx_{0\epsilon}$.
This formally proves that the final result for the partition function
$Z = \int d^Dx_0 \, (2\pi\beta)^{-{D\over 2}}
z^{(\rho)}(x_0) $
is gauge invariant, i.e. independent of the choice of the function $\rho$.
However,  the density $z^{(\rho)}(x_0)$ can in general 
depend on $\rho$ through total derivative terms which should  
vanish upon integration over the zero modes $x_0$.

One can derive the same result using BRST methods.
The action $S[x_0 +y]$ can be considered as a functional of both $x^\mu_0$ 
and $y^\mu(\tau)$ and is invariant under the shift (\ref{shift}).
To fix this gauge invariance one can introduce a constant ghost field
$\eta^\mu$ for this abelian shift symmetry. The corresponding BRST symmetry
transformation rules are
\bea
\delta x^\mu_0 &=& \eta^\mu \Lambda \nonumber\\[.5mm]
\delta y^\mu  &=& -\eta^\mu  \Lambda \nonumber\\[.5mm]
\delta \eta^\mu &=& 0 \ .
\eea
To gauge fix one must also introduce constant nonminimal 
fields $\bar \eta, \pi$ with the BRST rules
\bea
\delta \bar \eta_\mu &=& i \pi_\mu \Lambda \nonumber\\[.5mm]
\delta \pi_\mu &=& 0  \ .
\eea
Using the gauge fixing fermion 
\be
\Psi = \bar \eta_\mu \idt \rho(\tau ) y^\mu(\tau )
\label{gferm}
\ee
we obtain the gauge fixed action
\bea 
S_{gf} &=& S[x_0 +y] + {\delta\over \delta \Lambda}\Psi \nonumber\\[.5mm]
&=& S[x_0 +y] +  i \pi_\mu \idt \rho(\tau ) y^\mu(\tau )
-\bar \eta_\mu \eta^\mu
\eea 
where ${\delta\over \delta \Lambda}$ denotes a BRST variation
with the parameter $\Lambda$ removed form the left. 
The ghosts can be trivially integrated out while the integration over 
the auxiliary variable $\pi_\mu$ produces a delta function.
Thus, the BRST method reproduces the last line of (\ref{gfac}), however
it makes it easier to implement more general gauges 
(for example to get massive propagators), though they 
will not be considered  here.

To test the previous set-up, let us compute perturbatively in the 
proper time $\beta$ the partition function in eq. (\ref{A.pf}). 
From (\ref{gfac}) one can write it as
\bea
&& Z = \oint Dx\  e^{-S[x]} = \int {d^Dx_0 \over (2\pi\beta)^{D\over 2}}\,
 z^{(\rho)}(x_0) 
\nonumber\\[.5mm]
&& z^{(\rho)}(x_0) = \la e^{-S_{int}}\ra 
= \exp [ \la e^{-S_{int}}\ra_c -1]
\eea 
where ${(2 \pi \beta)^{-{D\over 2}}} $ is the well-known normalization
of the path integral measure, and the quantum average $\la e^{-S_{int}}\ra$ 
is obtained using the free propagators (i.e. from the action with $V=0$)
of the fields constrained by the gauge fixing. 
These quantum fields are given by 
$y^\mu(\tau)= x^\mu(\tau)-x^\mu_0$ with 
$x^\mu_0= \int_0^1 d\tau \rho(\tau) x^\mu(\tau)$ and their propagator reads 
\bea 
\la y^\mu(\tau) y^\nu(\sigma)\ra = - \beta \delta^{\mu\nu}
{\cal B}_{(\rho)}(\tau,\sigma) 
\label{A.prp}
\eea
where the Green function ${\cal B}_{(\rho)}(\tau,\sigma)$ 
has already been described in eq. (\ref{green}).

We now aim to compute eq. (\ref{A.pf}) to order $\beta^4$ by 
expanding the interacting action around the constant $x_0$
\bea 
S_{int}&=&
{1\over \beta} \idt \beta^2 V(x(\tau))\nonumber\\[.5mm]
&=& 
\beta \idt (
\underbrace{V}_{S_{4}} + 
\underbrace{y^\mu \partial_\mu V}_{S_5}
+ \underbrace{{1\over 2} y^\mu y^\nu \partial_\mu
\partial_\nu V}_{S_6} +\cdots + 
\underbrace{{1\over 6!}y^6 \partial^6 V}_{S_{10}}+ ...)
\eea
where for $S_{10}$ we used an obvious short-hand notation.
Here all vertices are evaluated at $x_0$ (i.e. one sets 
$x^\mu(\tau)= x_0^\mu+y^\mu(\tau)$ and expand 
the action around the constant $x_0^\mu$).
The vertices indicated by $S_n$ contribute effectively like $y^n$.
Considering that $S_4$ gives only disconnected contributions
and that only an even number of $y$ give  nonvanishing Wick contractions
we obtain
\bea
z^{(\rho)}(x_0) 
&=&
\la e^{-S_{int}}\ra \nonumber\\[.5mm]
&=&
\exp \Big[ 
-\la S_4+S_6+S_8+S_{10}\ra
+{1\over 2} \la S_5^2+ S_6^2+ 2S_5S_7\ra_c
\nonumber\\[.5mm]
&&+ O(\beta^5)\Big]
\eea 
and thus
\bea
Z={1\over (2 \pi \beta)^{D\over 2}}&&\!\!\int d^Dx_0
\exp \Big[ -\beta V +{\beta^2\over 2} M_1 \Box V
-{\beta^3\over 2} 
\Big ({1\over 4} M_2 \Box^2 V + M_4 (\partial_\mu V)^2\Big )\nonumber\\[.5mm]
&&
+\beta^4
\Big ({1\over 48} M_3 \Box^3 V 
+  {1\over 4} M_5 (\partial_\mu \partial_\nu  V)^2
+{1\over 2} M_6 (\partial_\mu  V) (\partial^\mu \Box V)
\Big )\nonumber\\[.5mm]
&&+ O(\beta^5)\Big]
\eea 
where the Wick contractions obtained using (\ref{A.prp})
produce the following integrals
\bea
M_1 &=& \idt  {\cal B}(\tau,\tau) =
- {1\over 12} + C_\rho  
\nonumber\\[.5mm]
M_2 &=& \idt  {\cal B}^2(\tau,\tau) =
{1\over 144} + C^2_\rho  - {1\over 6} C_\rho   + 4 C'_\rho
\nonumber\\[.5mm]
M_3 &=& \idt  {\cal B}^3(\tau,\tau) =
 - {1\over 1728} +C^3_\rho  -{1\over 4}  C^2_\rho  
+{1\over 48} C_\rho   -2  C'_\rho -8  C''_\rho
\nonumber\\[.5mm]
M_4 &=& \idts {\cal B}(\tau,\sigma) =
C_\rho  
\nonumber\\[.5mm]
M_5 &=& \idts {\cal B}^2(\tau,\sigma) =
{1\over 720} + C^2_\rho  +2 C'_\rho
\nonumber\\[.5mm]
M_6 &=& \idts {\cal B}(\tau,\sigma) {\cal B}(\sigma,\sigma)  =
C^2_\rho  -{1\over 12}  C_\rho  +2  C'_\rho
\nonumber
\eea
where on top of $C_\rho$ given in eq. (\ref{defd})
we have defined $C'_\rho = \int_0^1 d\tau F^2_\rho(\tau)$ and 
$C''_\rho = \int_0^1 d\tau F^3_\rho(\tau)$. Recall that the 
DBC propagator is obtained by
setting $\rho(\tau)=\delta(\tau)$ and the SI one 
by $\rho(\tau)=1$. In the SI case all these $C_\rho$'s vanish, while the
particular values of the integrals for the DBC propagators are as follows
\bea 
M_1 =  -{1\over 6}\, ,\ 
M_2 = {1\over 30}\, ,\ 
M_3 = -{1\over 140}\, ,\ 
M_4 = -{1\over 12}\, ,\ 
M_5 = {1\over 90}\, ,\ 
M_6 ={1\over 60} \ .
\nonumber
\eea

Let us now consider the partition function at order $\beta^4$
(we rename $x_0$ by $x$)
\bea
Z= {1\over (2 \pi \beta)^{D\over 2}} \int d^Dx 
&\Big [ &
1 -\beta V + {\beta^2\over 2} (V^2 + M_1 \Box V)\nonumber\\ 
&&+
{\beta^3\over 24} (-4V^3 -12 M_1 V \Box V - 3 M_2 \Box^2 V
-12 M_4 (\partial_\mu V)^2 ) \nonumber\\[.5mm]
&&+ 
\beta^4 \Big ({1\over 24} V^4 +{1\over 8} M_1^2 (\Box V)^2 
+{1\over 4} M_1 V^2\Box V +{1\over 2} M_4 V (\partial_\mu  V)^2 
\nonumber\\[.5mm]
&&+{1\over 48} M_3 \Box^3 V  
+{1\over 4} M_5 (\partial_\mu  \partial_\nu  V)^2 
+{1\over 2} M_6 (\partial_\mu   V) (\partial^\mu  \Box V) \Big)
\nonumber\\[.5mm]
&&+ O(\beta^5)
\Big] \ .
\eea
From this we deduce that
\bea
\Delta Z &\equiv& Z({\rm arbitrary}\ \rho)-Z(DBC)  =
{1\over (2 \pi \beta)^{D\over 2}} \int d^Dx 
\Big \{ 
{\beta^2\over 2} (M_1 + {1\over 6} ) \Box V
\nonumber\\[.5mm]
&&
- {\beta^3\over 24} \Big [ 3 (M_2 -{1\over 30}) \Box^2 V
+12 (M_1 + {1\over 6} ) 
\partial^\mu ( V\partial_\mu V) \Big ]
\nonumber\\[.5mm]
&&
+\beta^4 \Big [
{1\over 48}( M_3 +{1\over 140}) \Box^3 V 
+{1\over 8} (M_1^2 -{1\over 36} ) 
\partial^\mu ( \partial_\mu V \Box V) 
\nonumber\\[.5mm]
&&
+{1\over 4} (M_5 - {1\over 90} )
\partial^\mu ( \partial^\nu  V \partial_\mu  \partial_\nu  V)
+{1\over 8}  (M_2 -{1\over 30} ) 
\partial^\mu (  V \partial_\mu  \Box V)
\nonumber\\[.5mm]
&&
+{1\over 4} (M_1 + {1\over 6} ) 
\partial^\mu( V^2\partial_\mu  V) \Big ]
+ O(\beta^5)
\Big \} 
\eea
is a total derivative.
For potentials $V$ with derivatives 
vanishing sufficiently fast at infinity 
(a condition which is necessary for the validity of the perturbation
expansion in $\beta$) the total derivative can be dropped 
and all ways of factoring out the zero modes are equivalent.

\subsection{Conventions and identities}
\label{sub:conventions}
We use the following conventions for the curvatures
\bea
[\nabla_\alpha,\nabla_\beta]\, V^\mu &=& 
R_{\alpha\beta}{}^\mu{}_\nu V^\nu\nonumber\\[.5mm]
R_{\mu\nu} &=& R_{\lambda\mu}{}^\lambda{}_\nu\ . 
\eea 
The change of coordinates to Riemann normal coordinates is given by
\bea
y^\mu(x_0,\xi) = \xi^\mu - \sum_{n=2}^\infty \frac{1}{n!} 
   \nabla_{\a_3}\cdots\nabla_{\a_n}\Gamma^\mu_{\a_1\a_2}(x_0) \ 
   \xi^{\a_1} \cdots \xi^{\a_n}
\label{Arnc}
\eea
where the covariant derivatives act on lower indices only.

We have found it convenient to express all the terms in the 
order-$\beta$ correction to the chiral
anomaly (section~\ref{section:anomaly}) as combinations of the invariants
\bea 
K_0 &=& \varepsilon^{\alpha_1\cdots\alpha_4}\, 
R^{\mu\nu\r\s}\, R_{\r\s\alpha_1\alpha_2}\, R_{\mu\nu\alpha_3\alpha_4}
\nonumber\\[.5mm]
K_1 &=& \varepsilon^{\alpha_1\cdots\alpha_4}\, 
R^\s{}_{\mu\alpha_1}{}^\r\, R_{\s\alpha_2\nu\r}\, 
R^{\mu\nu}{}_{\alpha_3\alpha_4}\nonumber\\[.5mm]
K_2 &=& \varepsilon^{\alpha_1\cdots\alpha_4}\, 
R^\s{}_{\mu\alpha_1}{}^\r\, R_{\s\r\alpha_2\nu}\, 
R^{\mu\nu}{}_{\alpha_3\alpha_4}\nonumber\\[.5mm] 
K_3 &=& \varepsilon^{\alpha_1\cdots\alpha_4}\, 
R^\s{}_{\mu\alpha_1}{}^\r\, R_{\s\nu\r\alpha_2}\, 
R^{\mu\nu}{}_{\alpha_3\alpha_4}\ , 
\nonumber
\eea
where $K_1+K_2+K_3=0$ because of Bianchi identities.
By using Bianchi identities and integration by parts one can 
easily prove the following identities
\bea
\int d^4 x\,\varepsilon^{\alpha_1\cdots\alpha_4}\, 
\nabla_{\alpha} R^\alpha{}_{\s\alpha_1\alpha_2}\, \nabla_{\beta} 
R^{\beta\s}{}_{\alpha_3\alpha_4} &=& 
\int d^4 x\, \left(-{1\over 2}K_0 -2 K_1+4 K_2\right)\nonumber\\[.5mm]
\int d^4 x\,\varepsilon^{\alpha_1\cdots\alpha_4}\,
\Box R^{\beta\s}{}_{\alpha_1\alpha_2}\, R_{\beta\s\alpha_3\alpha_4} &=&
\int d^4 x\, 4\left(K_3- K_2\right)\nonumber\\[.5mm]
\int d^4 x\,\varepsilon^{\alpha_1\cdots\alpha_4}\,
R_{\r\s}\, R^\r_{\mu\alpha_1\alpha_2}\, R^{\s\mu}{}_{\alpha_3\alpha_4} &=&
\int d^4 x\, 2\left(K_2- K_1\right)\nonumber\\[.5mm]
\int d^4 x\,\varepsilon^{\alpha_1\cdots\alpha_4}\,
R_{\alpha_1}{}^\lambda\, R_{\mu\nu\lambda\alpha_2}\, 
R^{\mu\nu}{}_{\alpha_3\alpha_4} &=&
\int d^4 x\, \left({1\over 2}K_0- 2K_2\right)\nonumber\\[.5mm]
\int d^4 x\,\varepsilon^{\alpha_1\cdots\alpha_4}\,
R\, R_{\mu\nu\alpha_1\alpha_2}\, R^{\mu\nu}{}_{\alpha_3\alpha_4} &=&
\int d^4 x\, \left(2K_0- 8K_2\right)
\nonumber
\eea 
which have been used to cast the final results in a more compact form.
 
We now list few useful identities involving the propagators. 
Recalling the definition of $\Delta(\tau-\sigma)$ in (\ref{defd}),
we start by defining the quantities
\bea 
\begin{array}{ll}
{\displaystyle C_\r = \idtt \rho(\t_1)\ \rho(\t_2)\ \del(\t_1-\t_2)} &
\lx[C_{\rm DBC}=-{1\over 12}\rx]\\[.5mm]
{\displaystyle
D_\r = \idttt \r(\t_1)\ \r(\t_2)\ \r(\t_3)\ \del(\t_1-\t_2) 
\del(\t_1-\t_3)}& 
\lx[D_{\rm DBC}={1\over 144}\rx]\\[.5mm]
{\displaystyle
E_\r = \idtt \rho(\t_1)\ \rho(\t_2)\ \del{}^2(\t_1-\t_2)-{1\over 720}}&
\lx[E_{\rm DBC}={1\over 180}\rx]
\end{array}
\nonumber
\eea
which all vanish in the string inspired approach ($\rho(\tau)=1$).
For the $\rho$-dependent propagators we need 
\bea
&& \B(\t,\s) = \del(\t-\s)-\ids\!\!' \rho(\s')\,\del(\tau-\s') -
\ids\!\!' \rho(\s')\,\del(\s-\s') +C_\r                \nonumber\\[.5mm]
&& \B|_\t \equiv  \B(\t,\t) = C_\r -{1\over 12} -2\ids\!\!' \r(\s')\, 
\del(\t-\s')                                           \nonumber\\[.5mm]
&& \dB(\t,\s) = \ddel(\t-\s) - \ids\!\!' \r(\s')\, \ddel(\t-\s')
\nonumber\\[.5mm] 
&& \dB|_\t \equiv   \dB(\t,\t) = - \ids\!\!' \r(\s')\, \ddel(\t-\s') 
\nonumber\\[.5mm]
&& {d\over d\t}\B|_\t = 2\dB|_\t \label{rBd}\nonumber\\[2.2mm]
&& \dBd(\t,\s) = \ddeld(\t-\s)\nonumber\\[.5mm]
&& \F(\t,\s) = \ddel(\t-\s)-\ids\!\!' \rho(\s')\,\ddel(\tau-\s') +
\ids\!\!' \rho(\s')\,\ddel(\sigma-\s')\nonumber\\[.5mm]
&& \F|_\t \equiv  \F(\t,\t) = 0    \nonumber\\[1.5mm]
&& \dBd(\t,\s) +\dF(\t,\s) = 1-\r(\t)
\nonumber
\eea
and the identities
\bea
&&\int_0^1\!\!\! d\w\ \ddel(\t-\w)\, \ddel(\s-\w) =  - \del(\t-\s) 
\nonumber\\[.5mm]
&&\int_0^1\!\!\! d\w\ \del(\t-\w)\, \del(\s-\w) = -{1\over 6}\del^2(\t-\s)
-{1\over 36}\del(\t-\s) +{1\over 4320} \ .
\nonumber
\eea
These identities
have been used to the express all the worldline integrals, listed in 
the next appendix, in an economical form, namely as combinations of  
$C_\r,D_\r,E_\r$ and pure numbers.

\subsection{Integrals}
\label{sub:integrals}

We list here the worldline integrals needed in the main text,
evaluated with dimensional regularization whenever necessary.
We use the notations $\B\equiv \B(\t_1,\t_2)$, $\B|_\t \equiv \B(\t,\t)$,
and similarly for its derivatives.

\ \newline
\noindent In section 2.1 (partition function at 3 loops), we needed the 
following integrals
\bea
&& H_1 = \idt \B|_\t \ (\dBd+\Dgh)|_\t = C_\r -\frac{1}{12}        
\nonumber\\[.5mm]
&& H_2 = \idt \dB|_\t^2 = -C_\r                                    
\nonumber\\[.5mm]
&& H_3 = \idt \r(\t)\ \B|_\t = -C_\r -\frac{1}{12}                  
\nonumber\\[.5mm]
&& H_4 = \idt \B|_\t^2 \ (\dBd+\Dgh)|_\t = C_\r^2 -\frac{5}{18}C_\r 
     -\frac{2}{3}E_\r +\frac{1}{144}                               
\nonumber\\[.5mm]
&& H_5 = \idt \B|_\t \ \dB|_\t^2 = -C_\r^2 +\frac{1}{9}C_\r +D_\r 
     +\frac{1}{6}E_\r                                              
\nonumber\\[.5mm]
&& H_6 = \idtt \B|_1 \ \B|_2 \ ({\dBd}^2-\Dgh^2) =  -C_\r^2 
   +\frac{7}{18}C_\r +\frac{4}{3}E_\r -\frac{1}{144}               
\nonumber\\[.5mm]
&& H_7 = \idtt \B^2\ (\dBd+\Dgh)|_1 = C_\r^2 -\frac{1}{18}C_\r       
   -\frac{1}{3}E_\r +\frac{1}{720}                                 
\nonumber\\[.5mm]
&& H_8 = \idtt \dB|_1 \ \dBd\ \Bd|_2 \ \B = C_\r^2 
   -\frac{1}{12}C_\r -D_\r                                         
\nonumber\\[.5mm] 
&& H_9 = \idtt \dB|_1 \ \dB\ \Bd|_2 \ \Bd = C_\r^2 
   +\frac{1}{36}C_\r -D_\r +\frac{1}{6}E_\r                        
\nonumber\\[.5mm] 
&& H_{10} = \idtt \dB|_1 \ \dBd\ \B|_2 \ \Bd = C_\r^2       
   -\frac{1}{6}C_\r -D_\r -\frac{1}{2}E_\r                         
\nonumber\\[.5mm] 
&& H_{11} = \idtt \dB|_1 \ \dB\ \B\ (\dBd+\Dgh)|_2  = -C_\r^2  
   +\frac{1}{24}C_\r +\frac{1}{2}D_\r +\frac{1}{4}E_\r             
\nonumber\\[.5mm] 
&& H_{12} = \idtt \B|_1 \ \dB^2\ (\dBd+\Dgh)|_2 \ =  -C_\r^2       
   +\frac{7}{36}C_\r +D_\r +\frac{1}{6}E_\r -\frac{1}{144}         
\nonumber\\[.5mm] 
&& H_{13} = \idtt \B^2 \ ({\dBd}^2-\Dgh^2) = -C_\r^2 +\frac{1}{4}C_\r  
   +E_\r +\frac{1}{120}                                            
\nonumber\\[.5mm]
&& H_{14} = \idtt \dB^2\ {\Bd}^2 = C_\r^2 -2D_\r +E_\r 
   +\frac{1}{80}                                                   
\nonumber\\[.5mm] 
&& H_{15} = \idtt \dB\ \dBd\ \Bd\ \B = C_\r^2 -\frac{7}{72}C_\r 
   -D_\r -\frac{1}{3}E_\r -\frac{11}{1440}                         
\nonumber\\[.5mm] 
&& H_{16} = \idt \r(\t)\ \B|_\t^2 = -3C_\r^2 +\frac{1}{6}C_\r  
   +4D_\r +\frac{1}{144}                                           
\nonumber\\[.5mm]
&& H_{17} = \idtt \r(\t_1)\r(\t_2)\ \B^2 =C_\r^2 -2D_\r +E_\r 
   +\frac{1}{720}                                                  
\nonumber\\[.5mm]
&& H_{18} = \idtt \r(\t_2)\ \B^2\ (\dBd+\Dgh)|_1 = -C_\r^2 
   +\frac{1}{36}C_\r +D_\r +\frac{1}{6}E_\r +\frac{1}{720}         
\nonumber\\[.5mm]
&& H_{19} = \idtt \r(\t_2)\ \dB^2\ \B|_1 =  C_\r^2 
   -\frac{5}{36}C_\r -D_\r -\frac{5}{6}E_\r -\frac{1}{144}         
\nonumber\\[.5mm]
&& H_{20} = \idtt \r(\t_2)\ \B \ \dB^2\ \dB|_1 = C_\r^2 
   -\frac{1}{72}C_\r -\frac{3}{2}D_\r +\frac{5}{12}E_\r      \ .    
\nonumber
\eea

\ \newline
\noindent In the 4 loop calculation of section 2.2 
we made use of the following integrals
\bea
J_1 &=& 
\idtt \B\ \B|_2 \ (\dBd+\Dgh)|_2
=C_\r^2-{5\over 36}C_\r-{1\over 3}E_\r
\nonumber\\[.5mm]
J_2 &=&
\idtt \B\ (\dB{}^2)|_2
=-C_\r^2+{1\over 72}C_\r+{1\over 2}D_\r+{1\over 12}E_\r
\nonumber\\[.5mm]
J_3 &=&
\idtt \r(\t_2)\ \B\ \B|_2
=-2C_\r^2+2D_\r
\nonumber\\[.5mm]
J_4 &=&
\idtt \B{}^2\ (\dBd+\Dgh)|_2
=C_\r^2-{1\over 18}C_\r-{1\over 3}E_\r+{1\over 720}
\nonumber\\[.5mm]
J_5 &=&
\idtt \Bd{}^2\ \B|_2
=-C_\r^2+{7\over 36}C_\r+D_\r+{1\over 6}E_\r-{1\over 144}
\nonumber\\[.5mm]
J_6 &=&
\idtt \B\ \Bd\ \dB|_2
=-C_\r^2+{1\over 24}C_\r+{1\over 2}D_\r+{1\over 4}E_\r
\nonumber\\[.5mm]
J_7 &=&
\idtt \r(\t_2)\ \B{}^2
=-C_\r^2+{1\over 36}C_\r+D_\r+{1\over 6}E_\r+{1\over 720}
\nonumber\\[.5mm]
J_8 &=&
\idt \B{}^2|_\t
=C_\r^2-{5\over 18}C_\r-{2\over 3}E_\r+{1\over 144} \ .
\nonumber\\[.5mm]
J_9 &=& 
\idt \B|_\t 
= C_\r -{1\over 12}
\nonumber\\[.5mm]
J_{10} &=&
\idt \r(\t) \ \B|_\t 
= -C_\r-{1\over 12}
\nonumber\\[.5mm]
J_{11} &=& 
\int_{0}^{1} \!\!\! d\tau \
\B|_\tau  \ (\dBd + \Delta_{gh})|_\tau
= C_\r -{1\over 12}
\nonumber\\[.5mm]
J_{12} &=& 
\int_{0}^{1}  \!\!\! d\tau \
 \dB^2|_\tau =-C_\rho~.
\nonumber
\eea

\ \newline
\noindent Finally, the integrals needed in the calculation of the 
order-$\beta$ correction to the chiral anomaly in $D=4$ 
(section~\ref{section:anomaly}) are given by
\bea
I_1 &=&
\idtt \B\ \dBd\ \B|_2 ={5\over 36}C_\r +{1\over 3}E_\r -{1\over 144}
\nonumber
\\[.5mm]
I_2 &=&
\idtt \B|_1\ \dB\ \dB|_2 = -2I_3
\nonumber
\\[.5mm]
I_3 &=&
\idtt \dB|_1\ \B\ \dB|_2 = {1\over 36}C_\r +{1\over 6}E_\r
\nonumber
\\[.5mm]
I_4 &=&
\idtt \B|_1\ \dBd\ \B|_2 = 4I_3
\nonumber
\\[.5mm]
I_5 &=&
\idtt \dB\ \Bd\ \B =-{1\over 9}C_\r -{1\over 6}E_\r +{1\over 360}
\nonumber
\\[.5mm]
I_6 &=&
\idtt \B\ \B\ \dBd = -2 I_5 
\nonumber
\\[.5mm]
I_7 &=&
\int_{0}^{1}  \!\!\!  d\t_1 \! \int_{0}^{1}  \!\!\!  d\t_2 \! \int_{0}^{1}
\!\!\!  d\t_3 \
(\B\ \dFd)_{(1,2)}\ \B_{(2,3)}\ \dB_{(3,1)}  = I_1
\nonumber
\\[.5mm]
I_8 &=&
\int_{0}^{1}  \!\!\!  d\t_1 \! \int_{0}^{1}  \!\!\!  d\t_2 \! \int_{0}^{1}
\!\!\!  d\t_3 \
\B|_1\ \dF_{(1,2)}\ (\dB\ \Bd)_{(2,3)} 
=-{2\over 9}C_\r -{1\over 3}E_\r
\nonumber
\\[.5mm]
I_9 &=&
\int_{0}^{1}  \!\!\!  d\t_1 \! \int_{0}^{1}  \!\!\!  d\t_2 \! \int_{0}^{1}
\!\!\!  d\t_3 \
(\B\ \dF)_{(1,2)}\ \dB_{(2,3)}\ \dB_{(3,1)} 
= -{1\over 6}C_\r -{1\over 2}E_\r+{1\over 180}
\nonumber
\\[.5mm]
I_{10} &=&
\int_{0}^{1}  \!\!\!  d\t_1 \! \int_{0}^{1}  \!\!\!  d\t_2 \! \int_{0}^{1}
\!\!\!  d\t_3 \
(\Bd\ \dF)_{(1,2)}\ \B_{(2,3)}\ \dB_{(3,1)} 
=-{1\over 36}C_\r -{1\over 6}E_\r-{1\over 720}
\nonumber
\\[.5mm]
I_{11} &=&
\int_{0}^{1}  \!\!\!  d\t_1 \! \int_{0}^{1}  \!\!\!  d\t_2 \! \int_{0}^{1}
\!\!\!  d\t_3 \ (\dBd+\dF)|_1 \
\Bd_{(1,2)}\ \B_{(2,3)}\ \dB_{(3,1)} 
= {1\over 18}C_\r +{1\over 3}E_\r
\nonumber
\\[.5mm]
I_{12} &=&
\int_{0}^{1}  \!\!\!  d\t_1 \! \int_{0}^{1}  \!\!\!  d\t_2 \! \int_{0}^{1}
\!\!\!  d\t_3 \
\B|_1\ \dB_{(1,2)}\ \dBd_{(2,3)}\ \Bd_{(3,1)} 
= -{1\over 12}C_\r +{1\over 144}
\nonumber
\\[.5mm]
I_{13} &=&
\int_{0}^{1}  \!\!\!  d\t_1 \! \int_{0}^{1}  \!\!\!  d\t_2 \! \int_{0}^{1}
\!\!\!  d\t_3 \
\dB|_1\ \B_{(1,2)}\ \dBd_{(2,3)}\ \Bd_{(3,1)} 
=- {1\over 36}C_\r -{1\over 6}E_\r
\nonumber
\\[.5mm]
I_{14} &=&
\int_{0}^{1}  \!\!\!  d\t_1 \! \int_{0}^{1}  \!\!\!  d\t_2 \! \int_{0}^{1}
\!\!\!  d\t_3 \
(\dB\ \Bd)_{(1,2)}\ (\dB\ \Bd)_{(2,3)} 
= - {1\over 36}C_\r -{1\over 6}E_\r+{1\over 144}
\nonumber
\\[.5mm]
I_{15} &=&
\int_{0}^{1}  \!\!\!  d\t_1 \! \int_{0}^{1}  \!\!\!  d\t_2 \! \int_{0}^{1}
\!\!\!  d\t_3 \
(\F\ \Bd)_{(1,2)}\ \Bd_{(2,3)}\ \Bd_{(3,1)} 
= - {1\over 36}C_\r -{1\over 6}E_\r +{1\over 360}
\nonumber
\\[.5mm]
I_{16} &=&
\int_{0}^{1}  \!\!\!  d\t_1 \! \int_{0}^{1}  \!\!\!  d\t_2 \! \int_{0}^{1}
\!\!\!  d\t_3 \
(\F\ \B)_{(1,2)}\ \dBd_{(2,3)}\ \Bd_{(3,1)} 
=-{5\over 36}C_\r -{1\over 3}E_\r +{1\over 360}
\nonumber
\\[.5mm]
I_{17} &=&
\int_{0}^{1}  \!\!\!  d\t_1 \! \int_{0}^{1}  \!\!\!  d\t_2 \! \int_{0}^{1}
\!\!\!  d\t_3 \
(\F\ \dBd)_{(1,2)}\ \Bd_{(2,3)}\ \B_{(3,1)} 
=  {1\over 18}C_\r +{1\over 3}E_\r -{1\over 720}
\nonumber
\\[.5mm]
I_{18} &=&
\int_{0}^{1}  \!\!\!  d\t_1 \! \int_{0}^{1}  \!\!\!  d\t_2 \! \int_{0}^{1}
\!\!\!  d\t_3 \ \r(\t_1)\
\Bd_{(1,2)}\ \Bd_{(2,3)}\ \B_{(3,1)} = {1\over 36}C_\r +{1\over 6}E_\r 
+{1\over 720}\ .
\nonumber
\eea

\vfill\eject




\begin{thebibliography}{99}
\baselineskip=14pt

\bibitem{Schubert:2001he}
see for example: 
C.~Schubert,
Phys.\ Rept.\  {\bf 355} (2001) 73
[arXiv:hep-th/0101036],\\  and references therein.

\bibitem{Bastianelli:2002fv}
F.~Bastianelli and A.~Zirotti,
Nucl.\ Phys.\ B {\bf 642} (2002) 372
[arXiv:hep-th/0205182].

\bibitem{Bastianelli:2002qw}
F.~Bastianelli, O.~Corradini and A.~Zirotti,
Phys.\ Rev.\ D {\bf 67} (2003) 104009
[arXiv:hep-th/0211134].

\bibitem{Bastianelli:1998jm}
F.~Bastianelli, K.~Schalm and P.~van Nieuwenhuizen,
Phys.\ Rev.\ D {\bf 58} (1998) 044002
[arXiv:hep-th/9801105];
F.~Bastianelli and O.~Corradini,
Phys.\ Rev.\ D {\bf 60} (1999) 044014
[arXiv:hep-th/9810119].

\bibitem{DeBoer:1995hv}
J.~De Boer, B.~Peeters, K.~Skenderis and P.~Van Nieuwenhuizen,
Nucl.\ Phys.\ B {\bf 446} (1995) 211
[arXiv:hep-th/9504097];
Nucl.\ Phys.\ B {\bf 459} (1996) 631
[arXiv:hep-th/9509158].

\bibitem{Bastianelli:2000nm}
F.~Bastianelli, O.~Corradini and P.~van Nieuwenhuizen,
Phys.\ Lett.\ B {\bf 494} (2000) 161
[arXiv:hep-th/0008045].

\bibitem{Fliegner:1997rk}
D.~Fliegner, P.~Haberl, M.~G.~Schmidt and C.~Schubert,
Annals Phys.\  {\bf 264} (1998) 51
[arXiv:hep-th/9707189].

\bibitem{Schalm:1998ix}
K.~Schalm and P.~van Nieuwenhuizen,
Phys.\ Lett.\ B {\bf 446} (1999) 247
[arXiv:hep-th/9810115].

\bibitem{Friedan:1980jm}
D.~H.~Friedan,
Annals Phys.\  {\bf 163} (1985) 318.


\bibitem{Kleinert:2003zq}
H.~Kleinert and A.~Chervyakov,
Int.\ J.\ Mod.\ Phys.\ A {\bf 18} (2003) 5521
[arXiv:quant-ph/0301081].


\bibitem{Strassler:1992zr}
M.~J.~Strassler,
Nucl.\ Phys.\ B {\bf 385} (1992) 145
[arXiv:hep-ph/9205205].

\bibitem{Abbott:1980hw}
L.~F.~Abbott,
Nucl.\ Phys.\ B {\bf 185} (1981) 189.

\bibitem{DW} B.S. DeWitt, in ``{\em Relativity, Groups and Topology}''
(lectures at Les Houches 1963) ed. B. and C. DeWitt (Gordon Breach, NY, 1964);
``{\em Relativity, Groups and Topology II}'' (lectures at Les Houches 1983)
ed. B. DeWitt and R. Stora (North Holland, Amsterdam, 1984).

\bibitem{Alvarez-Gaume:1983at}
L.~Alvarez-Gaume,
Commun.\ Math.\ Phys.\  {\bf 90} (1983) 161;
L.~Alvarez-Gaume and E.~Witten,
Nucl.\ Phys.\ B {\bf 234} (1984) 269;
D.~Friedan and P.~Windey,
Nucl.\ Phys.\ B {\bf 235} (1984) 395.

\bibitem{Witten:df}
E.~Witten,
Nucl.\ Phys.\ B {\bf 202} (1982) 253.

\bibitem{Fradkin:1984pq}
E.~S.~Fradkin and A.~A.~Tseytlin,
Phys.\ Lett.\ B {\bf 158} (1985) 316;
Nucl.\ Phys.\ B {\bf 261} (1985) 1;
A.~A.~Tseytlin,
Phys.\ Lett.\ B {\bf 223} (1989) 165.

\bibitem{Howe:vm}
P.~S.~Howe, G.~Papadopoulos and K.~S.~Stelle,
Nucl.\ Phys.\ B {\bf 296} (1988) 26.

\bibitem{Bastianelli:1991be}
F.~Bastianelli,
Nucl.\ Phys.\ B {\bf 376} (1992) 113
[arXiv:hep-th/9112035];
F.~Bastianelli and P.~van Nieuwenhuizen,
Nucl.\ Phys.\ B {\bf 389} (1993) 53
[arXiv:hep-th/9208059].

\bibitem{Lee:vm}
T.~D.~Lee and C.~N.~Yang,
Phys.\ Rev.\  {\bf 128} (1962) 885; 
see also discussion at p. 62-63 in: 
E.~S.~Abers and B.~W.~Lee,
Phys.\ Rept.\  {\bf 9} (1973) 1.


\bibitem{Kleinert:1999aq}
H.~Kleinert and A.~Chervyakov,
Phys.\ Lett.\ B {\bf 464} (1999) 257
[arXiv:hep-th/9906156];
F.~Bastianelli, O.~Corradini and P.~van Nieuwenhuizen,
Phys.\ Lett.\ B {\bf 490} (2000) 154
[arXiv:hep-th/0007105].

\bibitem{Gilkey:iq}
P.~B.~Gilkey,
J.\ Diff.\ Geom.\  {\bf 10} (1975) 601.

\bibitem{Bastianelli:2000hi}
F.~Bastianelli, S.~Frolov and A.~A.~Tseytlin,
JHEP {\bf 0002} (2000) 013
[arXiv:hep-th/0001041].

\bibitem{Bastianelli:2000dw}
F.~Bastianelli and O.~Corradini,
Phys.\ Rev.\ D {\bf 63} (2001) 065005
[arXiv:hep-th/0010118].

\bibitem{Hatzinikitas:1997md}
A.~Hatzinikitas, K.~Schalm and P.~van Nieuwenhuizen,
Nucl.\ Phys.\ B {\bf 518} (1998) 424
[arXiv:hep-th/9711088].

\bibitem{Waldron:1995tq}
A.~K.~Waldron,
Phys.\ Rev.\ D {\bf 53} (1996) 5692
[arXiv:hep-th/9511148].

\bibitem{deBoer:2003cp}
J.~de Boer, K.~Schalm and J.~Wijnhout,
``General covariance of the non-Abelian DBI-action: Checks and balances,''
arXiv:hep-th/0310150.

\bibitem{Gopakumar:2003ns}
R.~Gopakumar,
Phys.\ Rev.\ D {\bf 70} (2004) 025009
[arXiv:hep-th/0308184].

\end{thebibliography}
\end{document}